\documentclass[11pt]{article}

\usepackage{latexsym,color,graphicx,amsmath,amssymb,subcaption}

\def\reals{{\mathbb R}}
\def\eps{{\varepsilon}}

\def\R{{\cal R}}
\def\T{{\cal T}}
\def\dd{{\sf dist}}
\def\ddv{{\sf dist_v}}

\renewcommand{\angle}{\measuredangle}

\newcommand{\adjustedaccent}[1]{%
  \mathchoice{}{}
    {\mbox{\raisebox{-.5ex}[0pt][0pt]{$\scriptstyle#1$}}}
    {\mbox{\raisebox{-.35ex}[0pt][0pt]{$\scriptscriptstyle#1$}}}
}

\newcommand\frownacc[1]{\overset{\adjustedaccent{\smallfrown}}{#1}}

\newtheorem{theorem}{Theorem}[section]
\newtheorem{lemma}[theorem]{Lemma}
\begin{document}

\title{Output sensitive algorithms for approximate incidences and their applications\footnote{
A preliminary version of this work appeared in Proc.\ 25th European Sympos. Algorithms (ESA), 2017, 5:1--5:13.}
}

\author{
Dror Aiger\thanks{%
  Google Inc.;
  email: {\tt aigerd@google.com}}
\and
Haim Kaplan\thanks{%
  School of Computer Science, Tel Aviv University, Tel~Aviv 69978, Israel;
  email: {\tt haimk@tau.ac.il}. Work by Haim Kaplan has been supported
by Grants  1161/2011 and 1367/2017 from the German-Israeli Science Foundation,
by Grants 1841-14 and 1595-19 from the Israel Science Foundation and by the
Blavatnik Research Fund in Computer Science at Tel Aviv University.}
\and
Micha Sharir\thanks{%
  School of Computer Science, Tel Aviv University, Tel~Aviv 69978, Israel;
  email: {\tt michas@tau.ac.il}. Work by Micha Sharir has been supported by
Grant 2012/229 from the U.S.-Israel Binational Science Foundation,
by Grant 892/13 from the Israel Science Foundation,
by the
Blavatnik Research Fund in Computer Science at Tel Aviv University, and by
the Hermann Minkowski--MINERVA Center for Geometry at Tel Aviv University.
Both Kapalan and Sharir have also been  supported by  the Israeli Centers for
Research Excellence (I-CORE) program (center no.~4/11).}
}

\maketitle

\begin{abstract}
An $\eps$-approximate incidence between a point and some geometric object (line, circle, plane, sphere)
occurs when the point and the object lie at distance at most $\eps$ from each other.
Given a set of points and a set of objects, computing the approximate incidences between
them is a major step in many database and web-based applications in computer vision and graphics,
including robust model fitting, approximate point pattern matching, and estimating the fundamental
matrix in epipolar (stereo) geometry.

In a typical approximate incidence problem of this sort, we are given a set $P$ of $m$ points
in two or three dimensions, a set $S$ of $n$ objects (lines, circles, planes, spheres),
and an error parameter $\eps>0$, and our goal is to report all pairs $(p,s)\in P\times S$
that lie at distance at most $\eps$ from one another. We present efficient output-sensitive
approximation algorithms for quite a few cases, including points and lines or circles in the plane,
and points and planes, spheres, lines, or circles in three dimensions. Several of these cases
arise in the applications mentioned above. Our algorithms report all pairs at distance $\le\eps$,
but may also report additional pairs, all of which are guaranteed to be
at distance at most $\alpha\eps$, for some problem-dependent constant $\alpha>1$.
Our algorithms are based on simple primal and dual grid decompositions and are easy to implement. We note that
(a) the use of duality, which leads to significant improvements in the overhead cost of the algorithms,
appears to be novel for this kind of problems; (b) the correct choice of duality in some of these problems
is fairly intricate and requires some care; and (c) the correctness and performance analysis of the
algorithms  (especially in the more advanced versions) is fairly non-trivial.
We analyze our algorithms and prove guaranteed upper bounds on their running time and on
the ``distortion'' parameter $\alpha$.

\end{abstract}

\section{Introduction} \label{sec:intro}

\paragraph{Approximate incidences.}
Given a finite point set $S_1$ and a finite set $S_2$ of geometric primitives (e.g., lines, planes,  circles, or spheres in $\reals^2$ or $\reals^3$),
and some $\eps >0$, we define the set of \emph{$\eps$-incidences} (also referred to as
\emph{$\eps$-approximate incidences}, or just \emph{approximate incidences}) between $S_1$ and $S_2$ to be
$$
I_\eps(S_1,S_2)=\{ (s_1,s_2) \mid s_1 \in S_1, s_2 \in S_2, \dd(s_1,s_2) \leq \eps\} ,
$$
where $\dd(s_1,s_2) = \inf \{\dd(s_1,y) \mid y\in s_2\}$
is the Euclidean distance between $s_1$ and $s_2$.
We are interested in efficient algorithms for
computing $I_\eps(S_1,S_2)$, ideally in time linear in $|S_1|+|S_2|+|I_\eps(S_1,S_2)|$.

Most of the classical work in discrete and computational geometry on this kind of problems is focused on
computing
exact incidences ($\eps=0$).
 The simplest,
and perhaps archetypal instance of this task is \emph{Hopcroft's problem}, where we want to determine whether
there exists at least one incidence between a set $S_1$ of $m$ points and a set $S_2$ of $n$ lines in the plane.
Solutions to this problem and its obvious generalizations run in time close to $m^{2/3}n^{2/3}+m+n$; see \cite{AE,PS:surv}.
The cases of more general families of curves or surfaces have received less attention. In principle,
 this problem is
a special case of \emph{batched range searching}, where the data set is $S_1$ and the ranges are the objects in $S_2$.
 These problems can be solved using standard range searching techniques,
as reviewed, e.g., in \cite{AE}, but the resulting running times, while subquadratic, are sometimes inferior to the best
known combinatorial bounds on the number of incidences (unlike the situation with Hopcroft's problem and its variants, where
the running time is similar to the incidence bound).
We note that a major difference between approximate incidences and exact
incidences is that the number of exact incidences is always asymptotically smaller
than $nm$, where $m=|S_1|$ and $n=|S_2|$,  whereas the number of approximate incidences could well be $nm$.

 In contrast, the notion of approximate
incidences, as we define here, has received less attention in the practical consideration, but it has many important applications which
we review below. We consider the problem of reporting all pairs in $I_\eps(S_1,S_2)$. Our algorithms, though, can also
estimate $|I_\eps(S_1,S_2)|$, rather than report its members, and do it faster when $|I_\eps(S_1,S_2)|$ is small.

The problem of finding approximate incidences  can also be viewed as a range searching problem. Specifically, we treat each member $s_2$
of $S_2$ as the range $s_2(\eps)=\{p\in \reals^d \mid \dd(p, s_2) \le \eps \}$.
Here $d$ is the dimension of the ambient space, which in this paper is $2$ or $3$.
By definition, $s_2(\eps)$ is the Minkowski sum of $s_2$ with a disk (ball in $\reals^3$) of radius $\eps$ (centered at the origin);
thus points become disks (in $\reals^2$) or balls (in $\reals^3$), lines become slabs (in $\reals^2$) or cylinders (in $\reals^3$), circles become
annuli (in $\reals^2$) or tori (in $\reals^3$), and so on. The goal now is to report all pairs $(s_1,s_2)\in S_1\times S_2$
such that $s_1\in s_2(\eps)$. As mentioned, the known algorithms for such tasks have a rather large overhead. For example,
when $S_1$ is a set of $m$ points and $S_2$ is a set of $n$ lines in the plane, i.e., the ranges $s_2(\eps)$
are fixed-width slabs, the best known algorithms for solving the problem have an overhead close to $m^{2/3}n^{2/3}$,
and there are matching lower bounds in certain models of computation.
The overhead is larger when the objects in $S_2$ are of more complex shapes
(e.g., arbitrary circles) or when we move to three (or higher) dimensions; see~\cite{AE}. In addition, these
algorithms, while interesting and sophisticated from a theoretical point of view, are a nightmare to implement in practice.

Instead, with the goal of obtaining algorithms that are really simple to implement (and therefore with good
performance in practice), and that run in time that is (nearly) linear in the input
and output sizes, we adopt the approach of using \emph{approximation schemes}, in which
we still report all the pairs $(s_1,s_2)$ that satisfy $\dd(s_1,s_2)\le\eps$, but are willing to report additional
pairs, provided that all pairs that we report satisfy $\dd(s_1,s_2)\le\alpha\eps$, for some constant problem-dependent parameter
$\alpha>1$. To be more precise, assuming that the test whether $\dd(s_1,s_2)\le\eps$ is cheap, we can filter
the reported pairs by such a test, and actually report only the pairs that pass it. The actual number of
pairs that we have to inspect will typically be larger than $|I_\eps(S_1,S_2)|$, but it will always be at
most $|I_{\alpha\eps}(S_1,S_2)|$ (and in practice considerably less than that),
and the hope is that the number of inspected pairs will not be much larger than those that we actually report.
(We expect it to be larger by only a constant factor, which depends on $\alpha$ and on the geometry of the setup under consideration.)

\paragraph{Our results.}

We present simple and efficient output-sensitive algorithms (in the above sense) for approximate-incidence reporting
problems between points and various simple geometric shapes, in two and three dimensions.

To calibrate the merits of our solutions, we first note that these approximate incidence reporting problems
can also be solved by naive grid-based algorithms, as follows. Consider, for example, the problem of reporting
approximate incidences between a set $S_1$ of $m$ points and a set $S_2$ of $n$ lines in the plane.
We assume that all the incidences that we seek  occur in the unit disk (ball in $\reals^3$).
We partition the unit disk by a uniform grid, each of whose cells is a square of side
length $\eps$. We store each point in $S_1$ in a bucket corresponding to the grid cell that contains it,
and, for each line $\ell\in S_2$, we report all the pairs involving $\ell$ and the points in the grid
cells that $\ell$ crosses, and in their neighboring cells. The running time is $O(m+n/\eps + k)$, where $k$
is the number of reported approximate incidences. Clearly, all pairs $(p,\ell)\in S_1\times S_2$ with $\dd(p,\ell)\le\eps$
are reported, and each reported pair $(p,\ell)$ satisfies $\dd(p,\ell)\le 2\sqrt{2}\eps$, as is easily checked.
If $n$ is much larger than $m$, we can use duality (where some care is needed to preserve point-line distances),
to map the points to lines and the lines to points, and thereby reduce the complexity to $O(n+m+\min\{m,n\}/\eps + k)$.
This method can also be applied in three dimensions, and yields the same time bounds as in the preceding
primal-only approach (duality is much trickier in these situations), namely, $O(m+n/\eps + k)$, when $S_2$ consists of one-dimensional objects
(e.g., lines or circles),
but the running time deteriorates to $O(m+n/\eps^2 + k)$ when $S_2$ consists of surfaces
(e.g., planes or spheres). In these latter cases (involving planes or \emph{congruent} spheres)
duality can be applied, to improve the time bound to $O(n+m+\min\{m,n\}/\eps^2 + k)$.

While superficially these simple solutions might look ideal, as they are linear in $m$, $n$, and $k$,
their dependence on $\eps$ is too naive and weak, and when $m$ and $n$ are large and $\eps$ small
(as is typically the case in practice), the algorithms are rather slow in practice.

In this paper we address this issue, and develop a series of ``primal-dual''
grid-based algorithms for several approximate incidence reporting problems, that are faster
than this naive scheme for suitable ranges of the parameters $m$, $n$, and $\eps$
(which cover most of the practical instances of these problems).
Specifically, we present the following results. In all of them, $S_1$ is a set of $m$ points,
contained in the unit ball in two or three dimensions, and $k$ is the number of points that we inspect;
the actual output size might be smaller.

\medskip
\noindent{\bf (a)}
In the plane, for a set $S_2$ of $n$ lines, all $k$ approximate incidences can be reported in time
$O\left(m+n+\sqrt{mn}/\sqrt{\eps}+k\right)$.
(The dependency of the complexity on $\eps$ is improved by a factor of $\sqrt{\eps}$ compared to
the naive scheme when $n$ and $m$ are comparable.). See Section \ref{sec:pl2d}.

\medskip
\noindent{\bf (b)}
In three dimensions, for a set $S_2$ of $n$ planes, all $k$ approximate incidences can be reported in time
$O\left(m+n+\sqrt{mn}/\eps+k\right)$.
(The dependency of the complexity on $\eps$ is improved by a factor of $\eps$ compared to the
naive scheme, when $n$ and $m$ are comparable.). See Section \ref{sec:pp3d}.

\medskip
\noindent{\bf (c)}
In the plane, for a set $S_2$ of $n$ congruent circles, all $k$ approximate incidences can be reported in time
$O\left(m+n+\sqrt{mn}/\sqrt{\eps}+k\right)$. See Section \ref{sec:circ2d}.

\medskip
\noindent{\bf (d)}
In the plane, for a set $S_2$ of $n$ arbitrary circles, all $k$ approximate incidences can be reported in time
$O\left(m+n+m^{1/3}n^{2/3}/\eps^{2/3}+k\right)$. See Section \ref{sec:arb_circ2d}.

\medskip
\noindent{\bf (e)}
In three dimensions, for a set $S_2$ of $n$ congruent spheres, all $k$ approximate incidences can be reported in time
$O\left((m+n)/\eps+k\right)$. See Section \ref{sec:cong_spheres3d}.

\medskip
\noindent{\bf (f)}
In three dimensions, for a set $S_2$ of $n$ lines, all $k$ approximate incidences can be reported in time
$O\left(m+n+m^{1/3}n^{2/3}/\eps^{2/3}+k\right)$. See Section \ref{sec:pl3d}.

\medskip
\noindent{\bf (g)}
In three dimensions, for a set $S_2$ of $n$ congruent circles, all $k$ approximate incidences can be reported in time
$$
O\left( (m+n)/\eps^{1/2} + m^{1/3}n^{2/3}/\eps^{7/6} + k \right) .
$$
See Section \ref{sec:pc3d}.

In Section \ref{sec:tri}, we use the algorithms in (e) and (g), to obtain an efficient algorithm
for finding triangles that are nearly congruent to a given triangle in a three-dimensional
point set. This is the first step in solving the approximate point pattern matching problem in $\reals^3$.
The  exact version of this problem (which is to report all triangles spanned by a set of $n$ points in 3-space which are congruent to a given triangle)
has been solved by Agarwal and Sharir~\cite{AS:cong}, in time close to $n^{5/3}$.

A comparison with the naive solutions sketched above clearly shows the superiority of our technique.
For example, for lines or congruent circles in the plane, assuming that $n \le m$, our algorithms
(in (a) and (c), respectively) are asymptotically faster than the naive method when
$\sqrt{mn/\eps} \le n/\eps$, that is, when $\eps \le n/m$, an assumption that holds in most practical applications.

To recap, one can obtain substantially better bounds than the naive ones. Our methods are based on grids and on duality---they
construct much coarser primal grids, and pass each subproblem, consisting of the points in a grid cell and of the objects
that pass through or near that cell, to a secondary dual stage, in which another coarse grid is constructed in a
suitably defined dual space. The output pairs are obtained from the cells of these secondary grids, and the gain is
in the overhead, as each primal or dual object crosses much fewer grid cells than in the naive solutions.
Although this primal-dual paradigm is fairly standard, its power in the approximate incidences context, as
considered here, has not been demonstrated before (to the best of our knowledge).
The analysis (and the particular duality one has to use) for some of the three-dimensional variants
is fairly challenging, but the algorithms all remain simple to describe and to implement.
We have actually implemented some of the algorithms and have experimented with them on several data sets.
This implementation is reported in Section \ref{sec:exp}.

\paragraph{Motivation and applications.}
Approximate incidence reporting and counting problems arise in several basic practical applications, in computer vision,
pattern recognition, and related areas. Three major applications of this sort are robust \emph{model fitting},
\emph{approximate point pattern matching} under rigid motions, and estimating the fundamental matrix in
(stereo) \emph{epipolar geometry}. All three problems share a common paradigm, which we first explain for model fitting.
In this problem, we are given a set $P$ of $n$ points, say in $\reals^3$ (typically, these are so-called
\emph{interest points}, extracted from some image or $3D$ sensors), and we want to fit objects (called \emph{models})
from some given family, such as lines, circles, planes, or spheres, so that each model passes near
(i.e., is approximately incident to) many points of $P$; the quality of the model is measured in
terms of the number of approximately incident points. The standard approach is to construct
(usually, by repeated random sampling) a sufficiently rich collection of candidate models. (For example,
for line models, one can simply sample pairs of points of $P$, and for each pair construct the line
passing through its points.) One then counts, for each candidate line, the number of approximately incident points
(for some specified error parameter $\eps>0$), and reports the models that have sufficiently many such points.

Similar reductions arise in the other problems.
In approximate point pattern matching, we are given two sets $A$, $B$ of points, and want to find rigid motions
that map sufficiently large subsets of $A$ to sets whose (unidirectional) Hausdorff distance to $B$ is at most $\eps$.
Here too we construct candidate rigid motions, and test the quality of each of them. For example, in the plane, we
sample pairs of points from $A$, and find, for each sampled pair, the pairs of points of $B$ that are nearly at the
same distance. For each such pair of pairs we construct a rigid motion that maps the first pair to near the other pair,
and then test the quality of each of these motions, namely, the number of points of $A$ that lie, after the motion,
near points of $B$. The first step can be reduced to approximate incidence counting involving circles (whose radii
correspond to the distance between the pairs of sampled points of $A$, and which are centered
at the points of $B$) and the points of $B$. In three dimensions, we need to sample triples of points of $A$,
and for each triple $a,b,c$, we need to find those triples of $B$ that span triangles that are nearly congruent
to $\Delta abc$ (because to determine a rigid motion in $\reals^3$ we need to specify how it maps three (noncollinear)
source points to three respective image points).
This step is described in detail in Section~\ref{sec:tri}.

In epipolar geometry, we have two stereo images $A$, $B$ of the same scene, and we want to estimate the fundamental
matrix $F$ that best matches $A$ to $B$, where a point $p\in A$ is (exactly) matched to a point $q\in B$ if
$p^TFq=0$. We construct a sample of candidate matrices, by repeatedly sampling $O(1)$ interest points from
both images, and test the quality of each matrix. To do so for a candidate matrix $F$, we left-multiply each point
$p\in A$ by $F$, interpret the resulting vectors $p^TF$, for $p\in A$, as lines, and count the approximate incidences
of each line with the points of $B$. If sufficiently many lines have sufficiently high counts, we regard $F$ as a
good fit and output it.

To recap, in each of these applications, and in other applications of a similar nature, we generate a random
sample of candidate models, motions, or matrices, and need to test the quality of each candidate. Approximate incidence reporting and
counting arises either in the generation step, or in the quality testing step, or in both. Improving the efficiency of
these steps is therefore a crucial ingredient of successful solutions for these problems. The standard approach, used
``all over'' in computer vision in practice, is the {\sf RANSAC} technique~\cite{R25,FB81}, which checks in brute force
each model against each point. Replacing it by efficient methods for  approximate incidence counting, which is our focus here,
can drastically improve the running time of these applications.

To support the claim that this is indeed the case in practice,
we have conducted, as already mentioned, preliminary experiments with some of our algorithms, tested them on real and random data,
and compared them with other existing methods. Roughly, they demonstrate that our approach is significantly
faster than the other approaches. Our experiments also support our feeling that the cost of reporting more
pairs than really needed (pairs that might be at most $\alpha\eps$ apart, rather than just $\eps$), is negligible
compared to the cost of the other steps (in themselves much more efficient than the competing techniques).
We leave the project of conducting a thorough experimental study for future work. While we will present the implementation that we have performed, the focus of this paper will be on developing the algorithms and
establishing their worst-case guarantees.

\paragraph{Related work.}
Model fitting and point pattern matching have been the focus of many studies, both theoretical and practical; see
for example~\cite{AK10,AMC08,Bra02,CM08,FM10,GIMV03,HS94,HU90,MAM14}.

We first note that in many of the common approaches used in practice (e.g., RANSAC for model fitting~\cite{R25,FB81}),
reporting or counting approximate incidences between models and points is done using brute force, examining every pair of a model and a point.
Some heuristic improvements have also been proposed (see, e.g., \cite{CM08} and the references therein).
A similar brute-force technique is commonly used for approximate point pattern matching too
(e.g., in the \emph{Alignment} method~\cite{HU90} and its many variants).

The use of (exact) geometric incidences in algorithms for
{\em exact} point pattern matching is well established; see, e.g., Brass~\cite{Bra02} for details.
Similar connections have also been used for the more practical problem of {\em approximate} point pattern matching.
Gavrilov et al.~\cite{GIMV03} gave efficient algorithms for approximate pattern matching in two and three dimensions
(where the entire sets $A$ and $B$ are to be matched),
that use algorithms for reporting approximate incidences. One of the main results in~\cite{GIMV03} is that in the
plane, all pairs of points at distance in $[(1-\eps)r,(1+\eps)r]$ can be reported in $O(n\sqrt{r/\eps})$ time, using a grid-based search.
(In a way, part of the study in this paper formalizes, extends, and improves this method.)

Aiger et al.~\cite{AMC08} proposed a method for point pattern matching in $\reals^3$, called $4$PCS (4-Points Congruent Sets),
which iterates over all pairs of coplanar quadruples of points, one from $A$ and one from $B$, that can be matched via an affine transformation,
and then tests the quality of each pair, focusing on pairs where the transformation is rigid.
This algorithm  does not use approximate incidences, and assumes the existence of coplanar tuples.

In a more recent work, Aiger and Kedem~\cite{AK10} describe another algorithm for computing approximate incidences of points
and circles, following a similar approach by Fonseca and Mount~\cite{FM10} for points and lines,
which is better than the one of~\cite{GIMV03} for $n=\Omega(1/\varepsilon^{3/2})$, and use this for
approximate point pattern matching. This algorithm has been used in
Mellado et al.~\cite{MAM14}, to reduce the running time of the
$4$PCS algorithm in~\cite{AMC08} to be asymptotically linear in $n$ and in the output size.

The method of \cite{AK10,FM10} provides an alternative approach to approximate incidence reporting,
for the cases of points and lines or congruent circles (the analysis in \cite{AK10} is rather sketchy, though).
This technique runs in $O(m+n+\log(1/\eps)/\eps^2+k)$ time. For the case of lines in the plane, the
scheme exploits the fact that we can approximate
(up to an error of $O(\eps)$) all lines in the plane that cross the unit disk, by $O(1/\eps^2)$ representative lines,
such that if a point in the unit disk is close to a representative line $\ell$, then it is also
close (up to some small negligible additive error) to all the lines in the input that $\ell$ represents (and vice versa). Assuming, for example,
that $m$ is constant, this alternative scheme is better than our new algorithm (for these restricted scenarios) when
$\sqrt{n}/\sqrt{\eps} \ge 1/\eps^2$, that is, when $n \ge 1/\eps^{3}$
(we ignore the factor $\log(1/\eps)$ in this calculation).
(This technique seems to be extendible to three dimensions, and to surfaces, but the formal details have not yet
been worked out, as far as we know.)

\section{Approximate incidences in planar point-line configurations} \label{sec:pl2d}

We consider the approximate incidences problem between a set $P$  of $m$ points in the unit disk $B$
in $\reals^2$, and a set $L$  of $n$ lines that cross $B$, with a given accuracy parameter $0<\eps\le 1/2$.

We approximate the distance $\dd(p,\ell)$ by
the vertical distance between $p \in P$ and $\ell \in L$, which we denote by $\ddv(p,\ell)$.
For this approximation to be good, the angle between $\ell$ and  the $x$-direction
should not be too large.
To ensure this, we partition $L$ into two subfamilies,
one consisting of the lines with positive slopes, and one of the lines with negative slopes.
We fix one subfamily, rotate the plane by $45^\circ$, and get the desired property. In what follows we assume that all the lines of $L$ are "nearly horizontal", in this sense.

\begin{figure}[!htb]
  \centering
     \includegraphics[trim = 40mm 10mm 40mm 20mm, clip, scale=0.3]{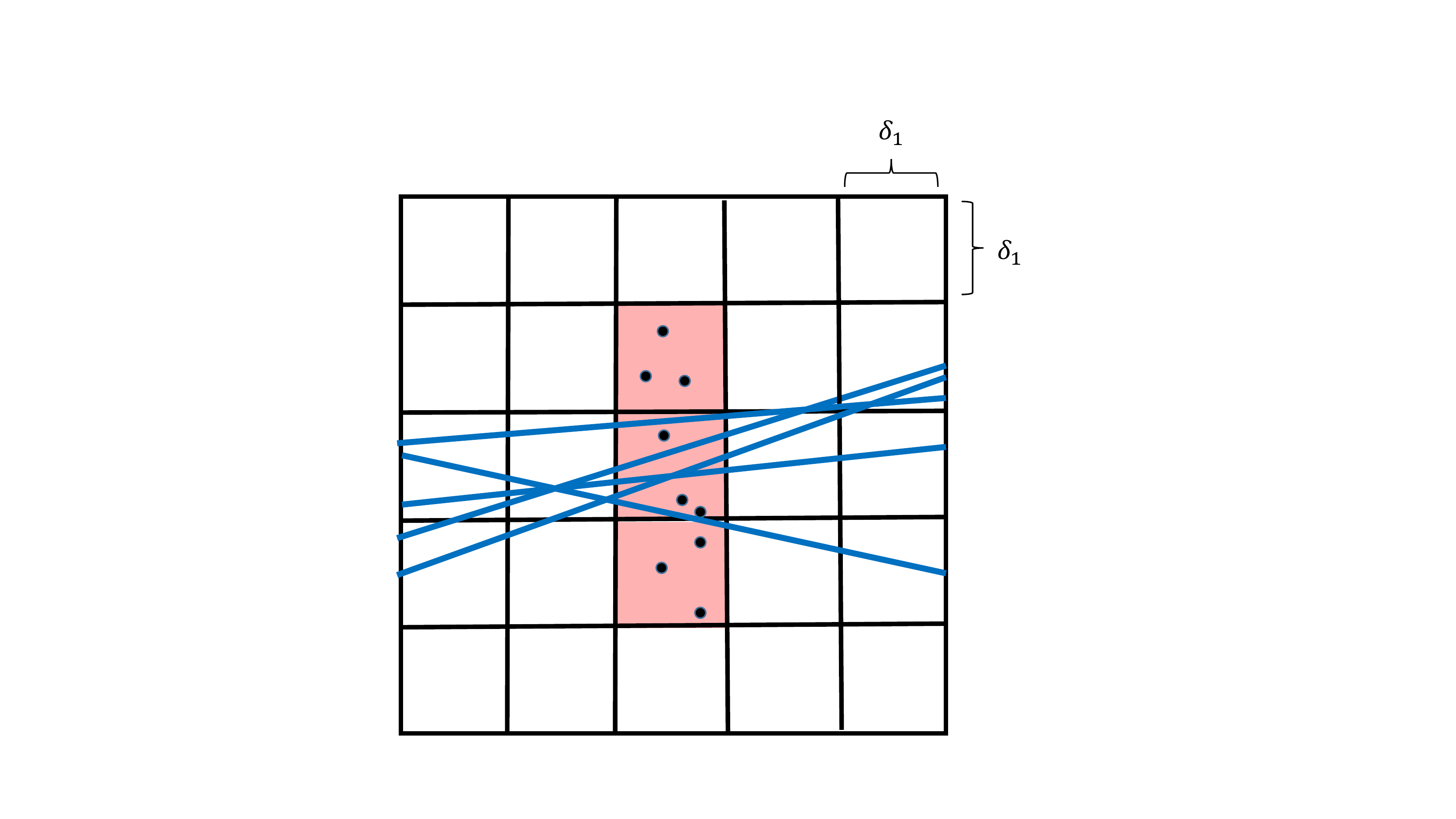}
    \caption{The partition of $S$ into subsquares, and the subproblem associated
 with the middle highlighted subsquare. \label{fig:lines}}
\end{figure}


Without loss of generality, we replace the unit disk $B$ by the unit square
$S=[0,1]^2$ (scaling down the plane by a factor of $2$), and apply the following two-stage partitioning procedure.
First we partition $S$ into
${1}/{\delta_1^2}$ pairwise openly disjoint smaller squares, each of side length
$\delta_1$, where $\delta_1$ is a parameter whose exact value will be set later. See Figure \ref{fig:lines}.  We ignore in what follows rounding issues and assume, for example, that
${1}/{\delta_1^2}$ is an integer.

Enumerate these squares as $S_1,S_2,\ldots,S_{1/\delta_1^2}$. For $i=1,\ldots,1/\delta_1^2$,
let $P_i$ denote the set of all points of $P$ that lie either in $S_i$ or in one
of the two squares that are directly above and below $S_i$ (if they exist), and let
$L_i$  be the set of all the lines of $L$ that cross $S_i$.
Put $m_i:=|P_i|$ and $n_i:=|L_i|$. We have $\sum_i m_i \le 3m$
and $\sum_i n_i \le 2n/\delta_1$, because each line of $L$ crosses at most $2/\delta_1$ squares $S_i$.

\begin{figure}[hbt]
\centering
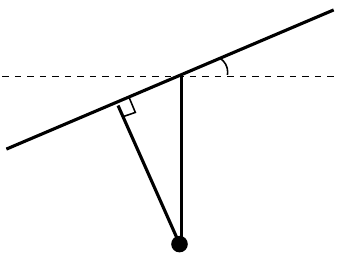
\caption{The relation between $\dd(p,\ell)$ and $\ddv(p,\ell)$.}
\label{fig:lines0}
\end{figure}

We now apply a duality transformation to each small square $S_i$ separately.
For notational simplicity, and without loss of generality, we may assume that
$S_i = [-\delta_1/2,\delta_1/2]^2$. (Technically, this means that we shift the cells by $\delta_1/2$ in both coordinate directions, so that the grid vertices now represent the centers of the cells.)
We map each point $p=(\xi,\eta)$ in $P_i$ to the line
$p^*:\; y=\xi x-\eta$, and each line $\ell:\;y=cx+d$ in $L_i$ to the point $\ell^*=(c,-d)$.
This duality preserves the vertical distance $\ddv$ between a point $p$
and a line $\ell$; that is, $\ddv(p,\ell) = \ddv(\ell^*,p^*)$.
Note that the slope condition ensures that
$\dd(p,\ell) \le \ddv(p,\ell) \le \sqrt{2}\dd(p,\ell)$. See Figure~\ref{fig:lines0}.

Let $\ell:\;y=cx+d$ be a line in $L_i$, that is, $\ell$ crosses $S_i$. By the slope
condition we have $-1\le c\le 1$ and
$-\delta_1\le d\le \delta_1$, so the dual point
$\ell^*$ lies in the rectangle $R:= [-1,1]\times [-\delta_1,\delta_1]$.
Each point $p=(\xi,\eta)\in P_i$ satisfies $-\delta_1/2\le \xi\le \delta_1/2$ and $-3\delta_1/2\le \eta\le 3\delta_1/2$
so the coefficients of the dual line $p^*:\;y=\xi x-\eta$ satisfy these respective inequalities.

We now partition $R$ into $1/\delta_2^2$ small rectangles, each of width $2\delta_2$
and height $2\delta_1\delta_2$, where $\delta_2$ is another parameter that we will shortly specify.
Each dual line $p^*$ crosses at most $2/\delta_2$ small rectangles.
To facilitate the following analysis, we choose $\delta_1$, $\delta_2$ so that they satisfy
$\delta_1\delta_2 = \eps$; we still have one degree of freedom in choosing them, which we will exploit later.
\begin{lemma} \label{lem:lines1-2d}
For each small rectangle $R'$, if $\ell^*$ is a dual point in $R'$ and $p^*$
is a dual line that crosses either $R'$ or one of the small rectangles directly above or below $R'$
(in the $y$-direction, if they exist), then the vertical distance $\ddv(\ell^*,p^*)$
(which is the same as $\ddv(p,\ell)$) is at most $5\delta_1\delta_2 = 5\eps$.
\end{lemma}
\noindent{\bf Proof.}
Indeed, if $p^*$ crosses a small rectangle $R''$, which is either $R'$ or one of the two adjacent rectangles, as above,
then, since the slope of $p^*$ is in $[-\delta_1/2,\delta_1/2]$, its maximum vertical deviation from $R''$ is at most
$2\delta_2 \cdot (\delta_1/2)=\delta_1\delta_2$.
Adding the heights $2\delta_1\delta_2$ of $R''$, and of $R'$ when $R''\ne R'$, the claim follows.
$\Box$

\begin{lemma} \label{lem:lines2-2d}
(a) Let $(p,\ell) \in P\times L$ be such that $\dd(p,\ell) \le\eps$.
Let $S_i$ be the small square containing $p$. If $\delta_1\ge \eps\sqrt{2}$, then $\ell$
must cross either $S_i$ or one of the two squares directly above and below $S_i$.
In other words, there exists a $j$ such that $(p,\ell)\in P_j\times L_j$.

\noindent (b) Continue to assume that $\dd(p,\ell) \le \eps$, let $i$ be such that
$(p,\ell)\in P_i\times L_i$, and let $R'$ be the dual small
rectangle (that arises in the dual processing of $S_i$) that contains $\ell^*$. Then
the dual line $p^*$ must cross either $R'$ or one of the two
small rectangles lying directly above and below $R'$ (in the $y$-direction, if they exist).
\end{lemma}
\noindent{\bf Proof.}
Both claims are obvious; in (a) we use the fact that $\ddv(p,\ell) \le \eps\sqrt{2}$,
and the assumption that $\eps\sqrt{2} \le \delta_1$;
see below how this is enforced.
In (b) we use the fact that $\ddv(p,\ell) = \ddv(\ell^*,p^*)$ and that the height of a small rectangle is
$2\delta_1\delta_2 = 2\eps > \eps\sqrt{2}$.
$\Box$

\paragraph{The algorithm.}
We first compute, for each point $p\in P$, the square $S_i$ it belongs to;
this can be done in $O(1)$ time, assuming a model of computation in which we can compute the floor function in constant time.
Similarly, we find, for each line $\ell\in L$, the squares that
it crosses, in $O(1/\delta_1)$ time. This gives us all the sets $P_i$, $L_i$, in overall
$O(m+n/\delta_1)$ time.

We then iterate over the small squares in the partition of $S$. For each such square $S_i$,
we construct the dual partitioning of the resulting dual rectangle $R$ into the smaller
rectangles $R'$. As above, we find, for each dual point $\ell^*$, for $\ell\in L_i$,
the small rectangle that contains it, and, for each dual line $p^*$, for $p\in P_i$,
the small rectangles that it crosses. This takes $O(n_i+m_i/\delta_2)$ time.

We now report, for each small rectangle $R'$, all the pairs $(p,\ell)\in P_i\times L_i$ for which
$\ell^*$ lies in $R'$ and $p^*$ crosses either $R'$ or one of the small rectangles
lying directly above or below $R'$ (if they exist). We repeat this over all small
squares $S_i$ and all respective small rectangles $R'$. Note that a pair $(p,\ell)$
may be reported more than once in this procedure, but its multiplicity is at most
some small absolute constant. The running time of this algorithm is
$$
O\left( m + \frac{n}{\delta_1} + \sum_{i=1}^{1/\delta_1^2} \left(n_i+\frac{m_i}{\delta_2}\right) + k \right) =
O\left(  \frac{n}{\delta_1}+\frac{m}{\delta_2}  + k \right) ,
$$
where $k$ is  the number of pairs that we report.
Lemma~\ref{lem:lines1-2d} guarantees that each reported pair is at distance $\le 5\eps$ and
 Lemma~\ref{lem:lines2-2d} guarantees that every pair $(p,\ell)$ at
distance at most $\eps$ is reported.

We optimize the running time by choosing $\delta_1$, $\delta_2$ to satisfy
$m/\delta_2 = n/\delta_1$ and $\delta_1\delta_2 = \eps$. That is, we want to choose
$ \delta_1 = \sqrt{{n\eps}/{m}}$ and
$\delta_2 = \sqrt{{m\eps}/{n}}$.
These choices are effective, provided that both $\delta_1$, $\delta_2$ are at most $1$,
for otherwise the primal partition or the dual partitions does not exist.
If $\delta_2 > 1$, that is, if $n < m\eps$, we simply choose $\delta_1=\eps$, and run only the primal
part of the algorithm, outputting all the pairs in $\bigcup_i P_i\times L_i$.
The cost is now $O(m+n/\eps+k) = O(m+k)$.
(This is the naive implementation, which is now efficient since $n$ is so small.)
If $\delta_1 > 1$, we pass directly to the dual plane, flip the roles of $P$ and $L$, and solve
the problem in the naive manner just described, at the cost of $O(n+k)$. Otherwise
(when both $\delta_1$ and $\delta_2$ are $\le 1$), the cost is
$O\left(\sqrt{mn}/\sqrt{\eps} + k \right)$. The cost of the algorithm is therefore always bounded by
$O\left( n+m+\sqrt{mn}/\sqrt{\eps} + k \right)$.

Recall also that in the proof of Lemma~\ref{lem:lines2-2d} we needed the inequality
$\eps\sqrt{2} \le \delta_1$. This will hold when $m\le n$ (and $\eps \le 1/2$, as we assume).
In the complementary case $m>n$, we simply flip the roles of points and lines (that is,
we start the analysis in the dual plane).

In conclusion, we have obtained the following main result of this section.
\begin{theorem} \label{th:lines-2d}
Let $P$ be a set of $m$ points in the unit disk $B$ in the plane,
let $L$ be a set of $n$ lines that cross $B$, and let $0<\eps\le 1/2$ be a prescribed parameter.
We can report all pairs $(p,\ell)\in P\times L$, for which $\dd(p,\ell)\le\eps$, in time
${\displaystyle O\left( n+m+{\sqrt{mn}}/{\sqrt{\eps}} + k \right)}$,
where $k$ is the actual number of pairs that we report; all pairs at distance at most $\eps$
are reported, and every reported pair lies at distance at most $5\eps$.
\end{theorem}

Another useful feature of the algorithm is that, rather than reporting all the pairs
that it produces, it can output a compact representation of them, as a union
of complete bipartite graphs $P_{\alpha} \times L_{\alpha}$. The number of such graphs is
$O\left(\frac{1}{\delta_1^2}\cdot \frac{1}{\delta_2^2} \right) = O(1/\eps^2)$,
and the sum of the cardinalities of their vertex sets is $O(m+n+\sqrt{mn}/\sqrt{\eps})$.
A similar feature holds for the algorithms in the forthcoming sections.

\section{Near neighbors in point-plane configurations} \label{sec:pp3d}

As a second application of the methodology illustrated in the preceding section, we
apply a similar approach in three dimensions. That is, given a set $P$ of $m$
points in the unit ball $B$ in $\reals^3$, a set $\Pi$ of $n$ planes crossing $B$,
and a prescribed error parameter $0< \eps\le 1/2$,
We solve the approximate incidences problem for $P$ and $\Pi$ with accuracy $\eps$.

We approximate the distance $\dd(p,\pi)$ by the $z$-vertical distance $\ddv(p,\pi)$
between $p$ and $\pi$. For this approximation to be good,  we partition $\Pi$ into $O(1)$ subfamilies,
such that, for each subfamily $\Pi'$ there exists a direction $u'$ such that the angle between
$u'$ and the normal of each plane of $\Pi'$ is at most $\pi/4$. We apply the construction to each subset
$\Pi'$ separately (with respect to all the points in $P$). When we apply it to a subfamily $\Pi'$,
we rotate the space such that $u'$ becomes the $z$-direction. In what follows, we fix one subfamily,
continue to denote it as $\Pi$, and assume that $u'$ is indeed the $z$-direction.

As in the two-dimensional case, we assume that all the points of $P$ are contained in
the unit cube $S=[0,1]^3$.

We apply a two-stage partitioning procedure analogous to the one of Section \ref{sec:pl2d}.
First we partition $S$ into $\frac{1}{\delta_1^3}$ pairwise openly disjoint smaller cubes,
each of side length $\delta_1$, where $\delta_1$ is a parameter whose exact value will be set later.

Consider one such small cube $S_i$, and assume that $S_i = [-\delta_1/2,\delta_1/2]^3$ (translate space by $-\delta_1/2$ in each axis).
Let $P_i$ denote the set of all points of $P$ that lie either in $S_i$ or in one
of the two cubes that lie directly above and below $S_i$ in the $z$-direction, (if they exist), and let
$\Pi_i$ be the set of all the planes of $\Pi$ that cross $S_i$.
Put $n_i:=|\Pi_i|$ and $m_i:=|P_i|$. We have $\sum_i m_i \le 3m$ and $\sum_i n_i = O(n/\delta_1^2)$,
because each plane of $\Pi$ crosses  $O(1/\delta_1^2)$ cubes $S_i$.

For each such $S_i$, we pass to the dual space, mapping each point $p=(\xi,\eta,\zeta)$ in $P_i$ to the plane
$p^*:\; z=\xi x+\eta y-\zeta$, and map each plane $\pi:\;z=ax+by+c$ in $\Pi_i$ to the point $\pi^*=(a,b,-c)$.
This duality preserves the vertical distance $\ddv$ between a point $p$
and a plane $\pi$; that is, $\ddv(p,\ell) = \ddv(\ell^*,p^*)$.
As in the planar case, the normal direction condition is easily seen to ensure that
$\dd(p,\ell) \le \ddv(p,\ell) \le \sqrt{2}\dd(p,\ell)$.

The normal direction condition also implies that, for each plane
$\pi:\;z=ax+by+c$ in the current subproblem,
$$
\frac{(-a,-b,1)\cdot (0,0,1)}{||(-a,-b,1)||}= \frac{1}{\sqrt{a^2+b^2+1}} \ge \cos(\pi/4) = \frac{1}{\sqrt{2}},
$$
so $a^2+b^2 \le 1$, and therefore $|a| \le 1$ and $|b| \le 1$.

Let $\pi:\;z=ax+by+c$ be a plane in $\Pi_i$. We then have $-1\le a\le 1$, $-1\le b\le 1$, and\footnote{%
  There exists a point $(x_1,y_1,z_1)\in \pi\cap S_i$, and then we have
  $\delta_1/2\le x_1\le \delta_1/2$, $-\delta_1/2\le y_1\le \delta_1/2$, $-\delta_1/2\le z_1\le \delta_1/2$, and $z_1 = ax_1+by_1 + c$.
  Thus $c=z_1-ax_1-by_1$, which, with $|a| \le 1$ and $|b| \le 1$, implies that $-3\delta_1/2\le c\le 3\delta_1/2$.}
$-3\delta_1/2\le c\le 3\delta_1/2$, so the dual point $\pi^*$ lies in the box $R:= [-1,1]^2\times [-3\delta_1/2,3\delta_1/2]$.
Each point $p=(\xi,\eta,\zeta)\in P_i$ satisfies $-\delta_1/2\le \xi\le \delta_1/2$, $\delta_1/2\le \eta \le \delta_1/2$, and
$-3\delta_1/2\le \zeta \le 3\delta_1/2$, so the coefficients of the dual plane $p^*:\; z=\xi x+\eta y-\zeta$
satisfy these respective inequalities.

We now partition $R$ into $1/\delta_2^3$ small boxes, each of $x$-range and $y$-range $2\delta_2$,
and of $z$-range $3\delta_1\delta_2$. Each dual plane $p^*$ crosses at most $O(1/\delta_2^2)$ small boxes.
We choose $\delta_1$, $\delta_2$ so that they satisfy $\delta_1 \geq \eps\sqrt{2}$ and $\delta_1\delta_2 = \eps$,
and prove lammas analogous to Lemma \ref{lem:lines1-2d} and Lemma \ref{lem:lines2-2d}. We omit both the statements
and the proofs, which are almost verbatim to those in Section \ref{sec:pl2d}. In the analog of Lemma \ref{lem:lines1-2d},
the constant $5$ has to be replaced by $7$, as is easily checked.

\smallskip
\noindent
{\bf The algorithm.}
We map each point
$p\in P$ to the cube $S_i$ containing it and each plane $\pi \in \Pi$ to the cubes that it crosses,
thereby obtaining all the sets $P_i$, $\Pi_i$. This takes $O(m+n/\delta_1^2)$ time.
We then iterate over the  cubes in the partition of $S$. For each such
cube $S_i$, we construct the dual partitioning of the resulting dual box
$R$ into the smaller boxes $R'$. As above, we find, for each dual point $\ell^*$,
the small box that contains it, and, for each dual plane $p^*$, the small boxes
that it crosses. This takes $O(n_i+m_i/\delta_2^2)$ time.

We now report, for each small box $R'$, all the pairs $(p,\ell)$ for which
$\ell^*$ lies in $R'$ and $p^*$ crosses either $R'$ or one of the small boxes
lying directly above or below $R'$ (in the third coordinate, if they exist). The overall running time is
$$
O\left(  \frac{n}{\delta_1^2}+\frac{m}{\delta_2^2}  + k \right) ,
$$
where $k$ is  the number of pairs that we report.

We optimize the running time by choosing $\delta_1$, $\delta_2$ to satisfy
$$
\frac{m}{\delta_2^2} = \frac{n}{\delta_1^2} , \quad\quad\text{and} \quad\quad \delta_1\delta_2 = \eps .
$$
That is, we choose
$$
\delta_1 = \left( \frac{n\eps^2}{m} \right)^{1/4} , \quad\quad\text{and} \quad\quad
\delta_2 = \left( \frac{m\eps^2}{n} \right)^{1/4} .
$$
As before, these choices make sense only when both $\delta_1$ and $\delta_2$ are at most $1$.
When one of them is larger than $1$, we proceed as in the two-dimensional case, performing either
only the primal stage or only the dual one, and obtain the cost $O(m+n+k)$. Thus, the total cost of the algorithm is
 $O\left(n+m+ \sqrt{mn}/\eps + k \right)$.
The requirement that $\delta_1 \geq \eps\sqrt{2}$ can be enforced as in the planar case.

In conclusion, we have obtained the following main result of this section.
\begin{theorem} \label{th:planes-3d}
Let $P$ be a set of $m$ points in the unit ball $B$ of $\reals^3$,
let $\Pi$ be a set of $n$ planes that cross $B$, and let $0<\eps\le 1/2$ be a prescribed parameter.
We can report all pairs $(p,\pi)\in P\times \Pi$ for which $\dd(p,\pi)\le\eps$, in time
$
O\left(n+m+ {\sqrt{mn}}/{\eps} + k \right) ,
$
where $k$ is the actual number of pairs that we report; all pairs at distance at most $\eps$
will be reported, and every reported pair lies at distance at most $7\eps$.
\end{theorem}

\section{Nearly congruent pairs in the plane} \label{sec:circ2d}

In this section we consider the following problem. We are given
 two point sets $P$, $Q$ in the plane, of respective sizes $m$ and $n$ (which would be the same set in some applications),
and we wish to report all pairs $(p,q)\in P\times Q$ such that $|pq|\in [r-\eps,r+\eps]$.
Here too we consider the approximation version, where we want all such pairs to be reported, and
want every reported pair to satisfy $|pq|\in [r-\alpha\eps,r+\alpha\eps]$, for a suitable absolute constant $\alpha$.
This problem is equivalent to an approximate incidences problem between $P$ and the set of congruent circles
 $C := \{c_q \mid q\in Q\}$ where  $c_q$ the circle of radius $r$ centered at a point $q$.
We assume that  $0< \upsilon \le r \le 1/2$ for some fixed positive  constant $\upsilon$.

In the following subsections we present two different solutions to the problem. The first solution,
inspired by a similar idea due to Indyk, Motwani, and Venkatasubramanian~\cite{Indyk:1999},
does not use duality. It is simple and elegant, but its major drawback is that it is not sensitive
to cases where $m$ and $n$ differ significantly. The second solution does use duality, and
is sensitive to such differences; it is closer to the preceding solutions for the point-line
and point-plane approximate incidences problems.

\subsection{Reporting all nearly congruent pairs in the plane I} \label{sec:pc2dI}

\begin{figure}[htbp]
  \begin{center}
    \includegraphics[trim = 40mm 75mm 0mm 5mm, clip, scale=0.5]{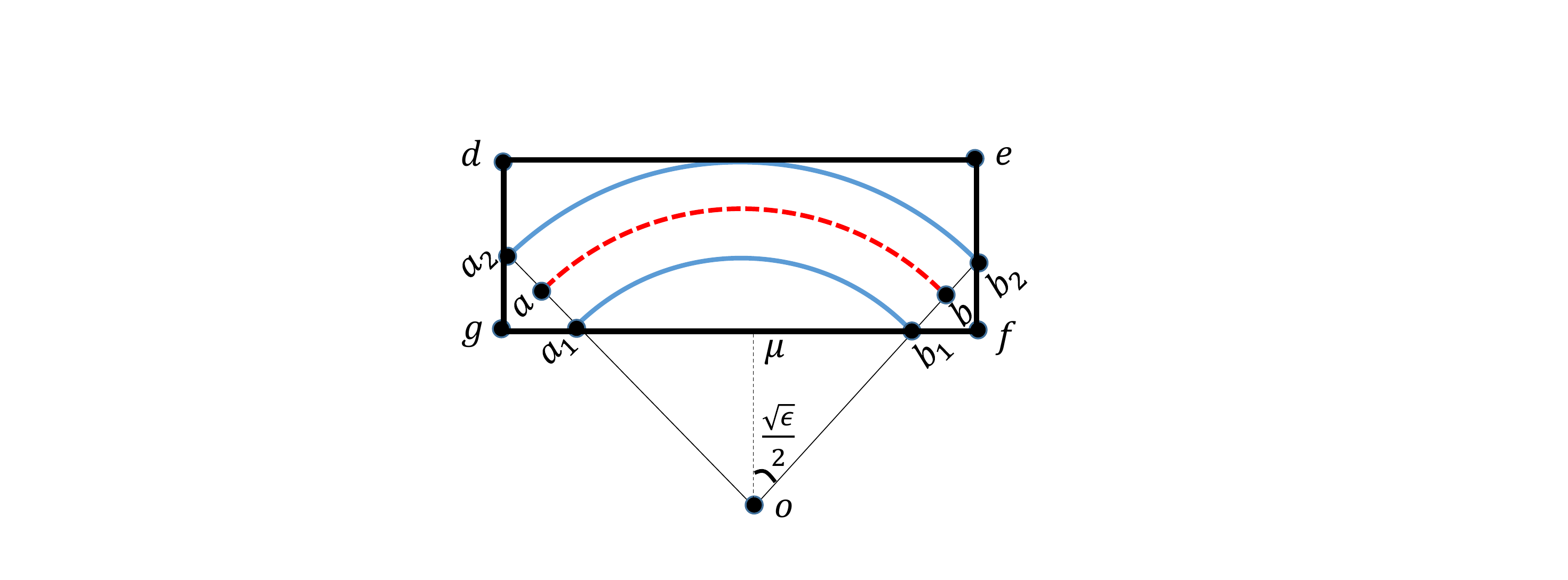}
  \end{center}
 \caption{The rectangle $R_{\overset{\frown}{ab}}$ that bounds
  the sector with central angle $\sqrt{\eps}$ of the annulus around the arc $\stackrel{\frown}{ab}$.
  \label{fig:circle3} }
\end{figure}

We take the circle $c_o$ of radius $r$ centered at the origin $o$, and partition it into $2\pi/\sqrt{\eps}$
equal canonical arcs, each with a central angle $\sqrt{\eps}$, delimited at the points on $c_o$ at orientations
$0,\sqrt{\eps},2\sqrt{\eps},\ldots$ (again, we ignore in what follows the routine rounding issues).
Consider one such arc $\frownacc{ab}$; see Figure \ref{fig:circle3}.
Let $A_o$ denote the annulus centered at the origin with inner radius $r-\eps$ and outer radius $r+\eps$.
Let $A_{\frownacc{ab}}$ be the portion of $A_o$ within the wedge $W_{ab}$ that defines the central angle of
$\frownacc{ab}$; that is, $W_{ab}$ is the wedge with $o$ as an apex, bounded by the rays $\vec{oa}$ and $\vec{ob}$.
Denote by $\frownacc{a_1b_1}$ and $\frownacc{a_2b_2}$ the respective inner and outer arcs that bound $A_{\frownacc{ab}}$.
Let $R_{\frownacc{ab}}$ be the smallest enclosing rectangle of $A_{\frownacc{ab}}$ whose longer side is parallel to
$ab$ (and to $a_1b_1$, $a_2b_2$); see Figure \ref{fig:circle3}.

The short edge, $ef$, of $R_{\frownacc{ab}}$ is of length
$$
2\eps + (r-\eps)-(r-\eps)\cos(\sqrt{\eps}/2)\le r+\eps-(r-\eps)(1-\eps/8)=\eps+r\eps/8 +\eps -\eps^2/8 \le 3\eps \ .
$$
The length of the large edge, $de$, of $R_{\frownacc{ab}}$ is
$$
2(r+\eps)\sin(\sqrt{\eps}/2) \le (r+\eps)\sqrt{\eps} \le \sqrt{\eps} ;
$$
In these derivations we use the inequalities $\cos x > 1-\frac12x^2$ and $\sin x < x$, for $x>0$,
and, in the very last inequality, also the fact that $ r\le 1/2$.
Note that these upper bounds on the side lengths of $R_{\frownacc{ab}}$ are tight up to a constant factor.

\begin{lemma} \label{lem:main-distR2}
(a) Let $q$ be a point at distance $\le\eps$ from $c_o$, so that the point of $c_o$ nearest to $q$ lies on $\frownacc{ab}$. Then
$q\in R_{\frownacc{ab}}$.

\smallskip
\noindent
(b) Let $R^*_{\frownacc{ab}}$ denote the homothetic copy of $R_{\frownacc{ab}}$ scaled by a factor of $3$
about its center. Then every point in $R^*_{\frownacc{ab}}$ is at distance $\le 5\eps$ from $c_o$.
\end{lemma}
\noindent{\bf Proof.}
(a) is trivial to prove because $q$ must lie in $A_{\frownacc{ab}}$.  For (b), we estimate the smallest and
largest distances from $o$ to points of $R^*_{\frownacc{ab}}$. The smallest distance is attained at the
midpoint $\mu^*$ of the longer edge of $R^*_{\frownacc{ab}}$ that is closer to $o$.

The distance of the midpoint $\mu$ of $gf$ (the longer edge of $R_{\frownacc{ab}}$
that is closer to $o$) from $o$
is equal to
$$
(r-\eps)\cos(\sqrt{\eps}/2) > (r-\eps)(1-\eps/8) > r - (1+r/8)\eps > r-2\eps .
$$
Since the width of $R_{\frownacc{ab}}$ is at most $3\eps$, the image $\mu^*$ of $\mu$
under this homothetic transformation
is closer to $o$ by at most $3\eps$, so the distance of $\mu^*$ from $o$ is at least $r-5\eps$.

The largest distance from $o$ to a point of $R^*_{\frownacc{ab}}$ is attained at
the images $d^*$ and $e^*$ of the respective vertices $d$ and $e$
of the longer edge of $R_{\frownacc{ab}}$ that is farther from $o$. To estimate
the distance from $o$ to $e^*$, say, we argue as follows. The image of the midpoint of $de$
is at distance at most $(r+\eps)+3\eps = r+4\eps$ from $o$, and half the length of the image of $de$ is at most
$3\sqrt{\eps}/2$. Hence the distance from $o$ to $e^*$ is at most
$$
\left( (r+4\eps)^2 + 9\eps/4 \right)^{1/2} =
\left( r^2 + (8r+9/4)\eps + 16\eps^2 \right)^{1/2} \leq r + \alpha\eps ,
$$
for any constant $\alpha$ satisfying $\alpha \geq 4$ and $\alpha \geq (8r+9/4)/2$, as is easily checked.
Since we assume that $r \leq 1/2$, we can take $\alpha=4$. This establishes (b).
$\Box$

Let $R_{\frownacc{ab}}(q)$ be the rectangle $R_{\frownacc{ab}}$ translated by the point (vector) $q$.
For each canonical arc $\frownacc{ab}$ of $c_o$, we consider all the rectangles $\{R_{\frownacc{ab}}(q) \mid q\in Q \}$,
and aim to find all pairs $(p,c_q)$, for $p\in P$, $q\in Q$, such that $p$ is contained in $R_{\frownacc{ab}}(q)$.
This is done as follows.

We rotate the plane such that each rectangle $R_{\frownacc{ab}}(q)$ becomes axis-parallel with
its long edge parallel to the $x$-axis (as depicted in Figure~\ref{fig:circle3}).
Clearly, in the rotated coordinate system, we can enclose all rectangles and points in a disk
centered at $o$ of radius slightly larger than $1$. Proceeding as in the previous sections,
we may assume that all our axis-parallel rectangles are contained in the unit square $S= [0,1]^2$.

We partition $S$ into a grid $G$ of isothetic copies of $R_{\frownacc{ab}}$, that is,
rectangles of size roughly $\sqrt{\eps}\times 3\eps$.
There are $O((1/\sqrt{\eps}) \cdot (1/\eps))=O(1/\eps^{3/2})$ such rectangles in $G$, and
each rectangle $R_{\frownacc{ab}}(q)$ intersects (the interiors of) at most four rectangles of $G$.
For each $q\in Q$, we report all the points of $P$ that lie in any of the four
corresponding rectangles of $G$.

The following lemma, combined with Lemma \ref{lem:main-distR2}, establishes the correctness of our scheme.
\begin{lemma}
We report all pairs $(p,q) \in P\times Q$ such that $p\in R_{\frownacc{ab}}(q)$.
Every pair $(p,q)$ that we report is such that $p$ is at distance at most $5\eps$ from $c_q$.
\end{lemma}

\noindent{\bf Proof.}
The first part is obvious. The second part follows from the observation that any grid cell that meets
$R_{\frownacc{ab}}(q)$ is fully contained in $R^*_{\frownacc{ab}}(q)$, which, combined with
Lemma~\ref{lem:main-distR2}(b), establishes the claim.
$\Box$

\smallskip
It takes $O(m)$ time to assign each point of $P$ to the cell of $G$ that contains it.
It then takes $O(n+k_{ab})$ time to find and report all the $k_{ab}$ pairs $(p,q)$ such that $p$
lies in one of the four grid cells that $R_{\frownacc{ab}}(q)$ overlaps.
Thus the total running time per arc $\frownacc{ab}$ is $O(m+n+k_{ab})$.
Adding up these bounds, over all $O(1/\sqrt{\eps})$ arcs $\frownacc{ab}$ we get that the total running time is $O((m+n)/\sqrt{\eps}+k)$,
where $k$ is the total number of reported pairs.,
Clearly, every pair $(p,q)$, where $p$ is at distance $\leq\eps$ from $c_q$, is reported, and we
 report each pair
$(p,q)$ only a constant number of times.

\begin{theorem} \label{th:circles1-2d}
Let $P$ and $Q$  be two sets of $m$  and $n$ points, respectively, in the unit disk $B$, and let
$0<\upsilon \le r\le 1/2$ for some fixed constant $\upsilon$.
We can report all pairs $(p,q)\in P\times Q$ for which $\dd(p,q)\in [r-\eps,r+\eps]$, in time
$$
O\left( \frac{m+n}{\sqrt{\eps}} + k \right) ,
$$
where $k$ is the actual number of pairs that we report; all pairs at distance in $[r-\eps,r+\eps]$
will be reported, and every reported pair lies at distance in $[r-5\eps,r+5\eps]$.
\end{theorem}

\subsection{Reporting all nearly congruent pairs in the plane II} \label{sec:pc2dII}

We next present an alternative approach to the problem considered in the preceding subsection.
Let $P$, $Q$, $m$, $n$, $r$, and $\eps$ be as above.
Again, we may assume that $P$ and $Q$ are bounded in the unit square $S=[0,1]^2$.

We apply a two-stage partitioning procedure, similar to the one given for the cases of lines and planes.
We fix two real positive parameters $\delta_1$, $\delta_2$, whose values will be set later.
First we partition $S$ into ${1}/{\delta_1^2}$ pairwise openly disjoint smaller
squares, each of side length $\delta_1$.
Enumerate these squares as $S_1,\ldots,S_{1/\delta_1^2}$.
Let $\hat{S}_i$ denote the union of $S_i$ and the (at most) eight squares adjacent to $S_i$.
Let $P_i$ denote the set of all points of $P$ that lie in $\hat{S}_i$, and let $C_i$
denote the set of all the circles $c_q\in C$  that cross $S_i$.
Put $m_i:=|P_i|$ and $n_i:=|C_i|$, for $i=1,\ldots,1/\delta_1^2$. We have
$\sum_i m_i = O(m)$, and $\sum_i n_i = O(n/\delta_1)$.

Fix a small square $S_i$. To find all the $\eps$-near pairs among points in $P_i$ and circles in $C_i$,
we pass to the dual plane, where (i) we map each point $p\in P_i$ to the circle $c_p$ of radius $r$ centered
at $p$, and (ii) we map each circle $c_q\in C_i$ to its center $q$ (so now the elements of $Q$ become
points and those of $P$ become circles). The distance between $q$ and $c_p$
is the same as the distance between $p$ and $c_q$.

Let $c_q$ be a circle in $C_i$. Clearly, $q$ has to lie in the Minkowski sum $K_i$ of $S_i$ and  the
circle of radius $r$ centered at the origin. As is easily checked, $K_i$ is contained in the annulus
that is centered at the center $o_i$ of $S_i$ and has radii $r\pm \delta_1/\sqrt{2}$ (note that $\delta_1/\sqrt{2}$
is half the diameter of $S_i$). (We assume that $\delta_1 < r$.) To simplify the notation, denote this annulus also as $K_i$; we
will use this annulus instead of the Minkowski sum in what follows.

Passing to polar coordinates $(\rho,\theta)$ about $o_i$, we get that $K_i$ becomes the rectangle
$$
R = [r-\delta_1/\sqrt{2}, r+\delta_1/\sqrt{2}] \times [0,2\pi] .
$$
We partition $R$ into $1/\delta_2^2$ small (polar) rectangles, each of width ($\rho$-range) $\sqrt{2}\delta_1\delta_2$
and height ($\theta$-range) $2\pi\delta_2$; in the standard coordinate frame, each small
rectangle is a sector of some (narrower) annulus centered at $o_i$, with the above width and angle.
Each dual circle $c_p$ crosses at most $O(1/\delta_2)$ small rectangles (that is, annulus sectors) of this grid.
This easily follows from the fact that the circle
$c_p$ is the graph of a well-defined function $r=f_p(\theta)$, of constant complexity,
in our polar coordinate frame.

To facilitate the following analysis, we will choose $\delta_1\ll r$ (recall that  $\upsilon \le r$ for some fixed constant $\upsilon$),
$\delta_2 \le \sqrt{\eps/r}$, and $\delta_1\delta_2 = \sqrt{2}\eps$; see below for the way
in which we ensure that these constraints hold.
The latter choice makes the $\rho$-range of each small
polar rectangle in the decomposition of $R$ equal to $\sqrt{2}\delta_1\delta_2 = 2\eps$.

\begin{lemma} \label{lem:circle2}
Let $R'$ be a small polar rectangle in the decomposition in the dual problem of $S_i$. Let $q$ be a dual point in $R'$, and let
$c_p$ be a dual circle that crosses $R'$ or one of its two adjacent rectangles with the same $\theta$-range.
Then $r-\alpha\eps \le |pq| \le r+\alpha\eps$ for some suitable absolute constant $\alpha$.
\end{lemma}

\begin{figure}[htbp]
  \begin{center}
    \includegraphics[trim = 0mm 10mm 0mm 0mm, clip, scale=0.4]{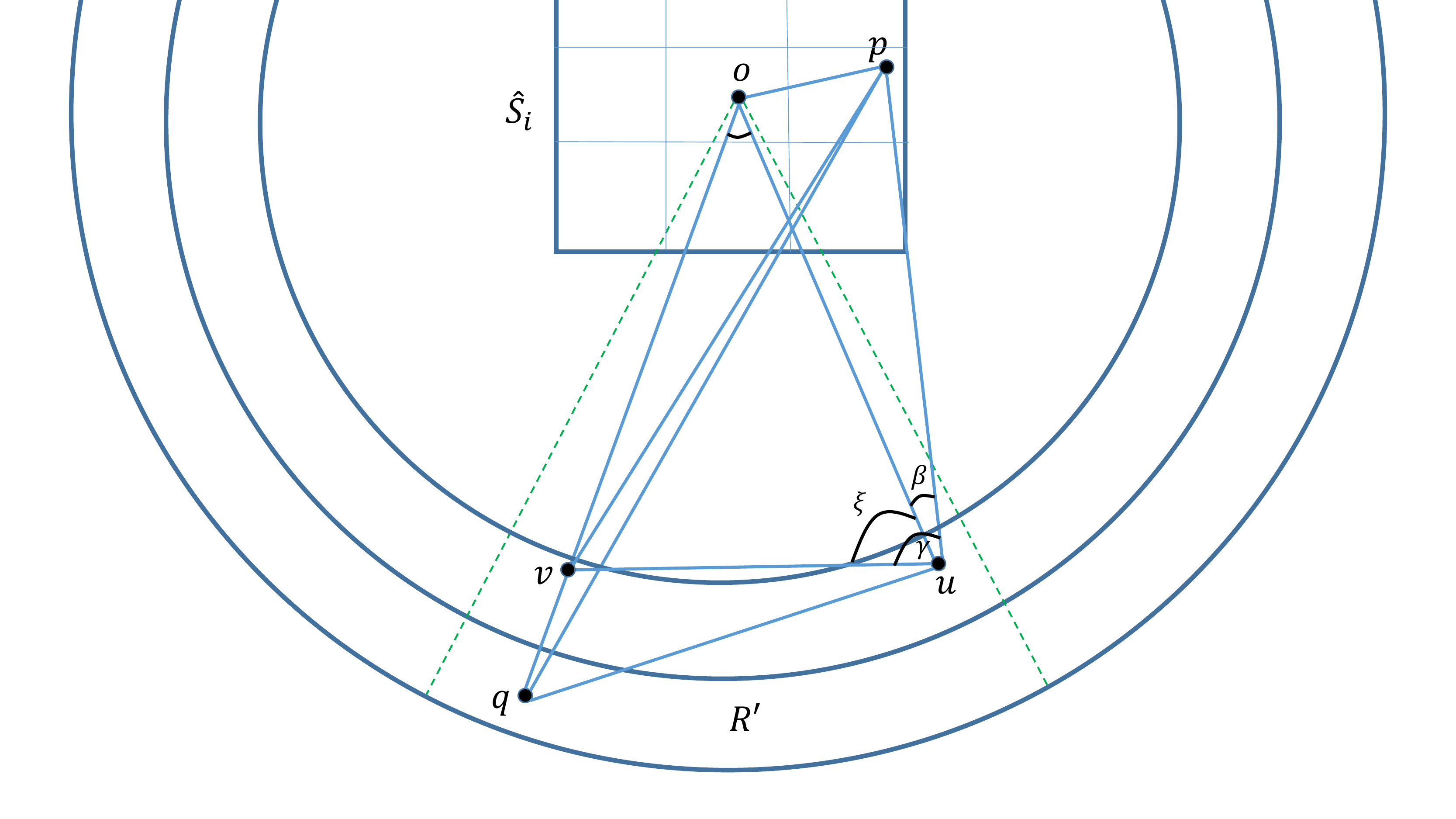}
  \end{center}
  \caption{\small An illustration of the proof of Lemma \ref{lem:circle2}. }
  \label{fig:circles2}
\end{figure}

\noindent{\bf Proof.}
Let $u$ be a point in the intersection of $c_p$ with $R'$ or with one of its adjacent rectangles
with the same $\theta$-range, and let $o=o_i$ be the center of $S_i$; see Figure \ref{fig:circles2}.
We know that (i) $|pu|=r$, (ii) $|op| \le 3\delta_1/\sqrt{2}$,
(iii) $\big| |ou|-r \big| \le \delta_1/\sqrt{2}$, and
(iv) $\big| |oq|-|ou| \big| \le 2\sqrt{2}\delta_1\delta_2 = 4\eps$.
We want to show that $r-\alpha\eps\le |pq| \le r+\alpha\eps$, for some absolute constant $\alpha$.

Let $v$ be the point on $oq$ satisfying $|ov|=|ou|$. By (iv), we have $|qv|\le 4\eps$.
It therefore suffices to show that $\big| |pv|-r \big| \le c\eps$, for a suitable constant $c$.

In the isosceles triangle $\Delta_{uov}$, the angle at $o$ is at most $2\pi\delta_2$, so its base
$uv$ is of length
$$
|uv| \le 2|ou|\sin \pi\delta_2 \le |ou|\cdot 2\pi\delta_2 \le \left(r+\delta_1/\sqrt{2}\right) \cdot 2\pi\delta_2 \le 10r\delta_2 ,
$$
which can be assumed in view of (iii) and the assumption that $\delta_1 \ll r$. Moreover, since $\Delta_{uov}$ is isosceles we have:
\begin{equation} \label{eqn:xi}
\xi = \angle ouv = \frac{\pi}{2} - \frac{\angle uov}{2} \ge \frac{\pi}{2} - \pi\delta_2 .
\end{equation}
Consider next the triangle $\Delta oup$ and its angle $\beta = \angle puo$. By the Law of Sines, we have
$$
\frac{|op|}{\sin\beta} = \frac{|pu|}{\sin \angle pou} \ge |pu| = r .
$$
Hence, by (ii)
$$
\sin\beta \le \frac{|op|}{r} \le \frac{3\delta_1}{r\sqrt{2}} ,
$$
so we may conclude that $\beta \le \frac{3\delta_1}{r}$, again under the assumption that $\delta_1 \ll r$.

Consider now the triangle $\Delta puv$, and let $\gamma$ denote its angle $\angle puv$.
Regardless of how the two triangles $\Delta puo$ and $\Delta ouv$ are juxtapositioned, we have
$$
\xi-\beta \le \gamma \le \xi+\beta.
$$
Subtracting this inequality from $\pi/2 $ we get
$$
\frac{\pi}{2} - (\xi-\beta) \ge \frac{\pi}{2} -\gamma \ge \frac{\pi}{2} - (\xi+\beta).
$$
Since $\xi \le \pi/2$ the right hand side is at least $-\beta$, and by Equation (\ref{eqn:xi})
the left hand side is at most $\beta + \pi \delta_2/2$.

Combining this with our conclusion above that $\beta \le \frac{3\delta_1}{r}$ we get that
$$
\left| \frac{\pi}{2} - \gamma \right| \le \beta + \pi\delta_2 \le
\frac{3\delta_1}{r} + \pi\delta_2 .
$$
Hence, by the Law of Cosines,
$$
|pv|^2 = |pu|^2 + |uv|^2 -2|pu| |uv|\cos\gamma
= r^2 + |uv|^2 - 2r|uv|\sin\left( \pi/2 - \gamma \right) .
$$
Write the right-hand side as $r^2(1+x)$, where
\begin{align*}
|x| & = \frac{1}{r^2}  \big| |uv|^2 - 2r|uv|\sin\left( \pi/2 - \gamma \right) \big| \\
& \le \frac{1}{r^2} \left( |uv|^2 + 2r|uv|\left( 3\delta_1/r + \pi\delta_2 \right) \right)
= O\left( \delta_2^2 + \frac{\delta_1\delta_2}{r} \right) .
\end{align*}
We thus have $|pv| = r(1+x)^{1/2}$, and $1-|x| \le (1+x)^{1/2} \le 1+\frac12 |x|$
(where the left inequality holds for $|x| < 1$, which we may assume to be the case).
In other words,
$$
\big| |pv|-r \big| \le r|x| = O\left( r\delta_2^2 + \delta_1\delta_2 \right) ,
$$
which, by the assumptions we have made, is $O(\eps)$, as asserted.
$\Box$

\begin{lemma} \label{lem:circle1}
(a) Let $(p,q) \in P\times Q$ be such that $r-\eps\le |pq| \le r+\eps$. Let $S_i$ be the small square containing $p$.
Then $c_q$ must cross either $S_i$ or one of its adjacent squares. So there must be a (unique) index $j$ such that $(p,c_q)\in P_j\times C_j$.

\smallskip
\noindent
(b) Let $j$ be the index for which $(p,c_q)\in P_j\times C_j$, and let $R'$ be the dual small
polar rectangle (that arises in the dual processing of $S_j$) that contains $q$. Then
the dual circle $c_p$ must cross either $R'$ or one of the two
small rectangles lying directly above and below $R'$ (in the $\rho$-direction, if they exist).
\end{lemma}
\noindent{\bf Proof.}
The proof of part (a) is trivial: Since the distance between $c_q$ and $p$ is at at most
$\eps = \delta_1\delta_2/\sqrt{2} \le \delta_1$ (the latter inequality holds since
$\delta_2\le 1$, by construction), $c_q$ must cross a square $S_j$ adjacent to $S_i$.

\begin{figure}[htbp]
  \begin{center}
    \includegraphics[trim = 0mm 10mm 0mm 0mm, clip, scale=0.4]{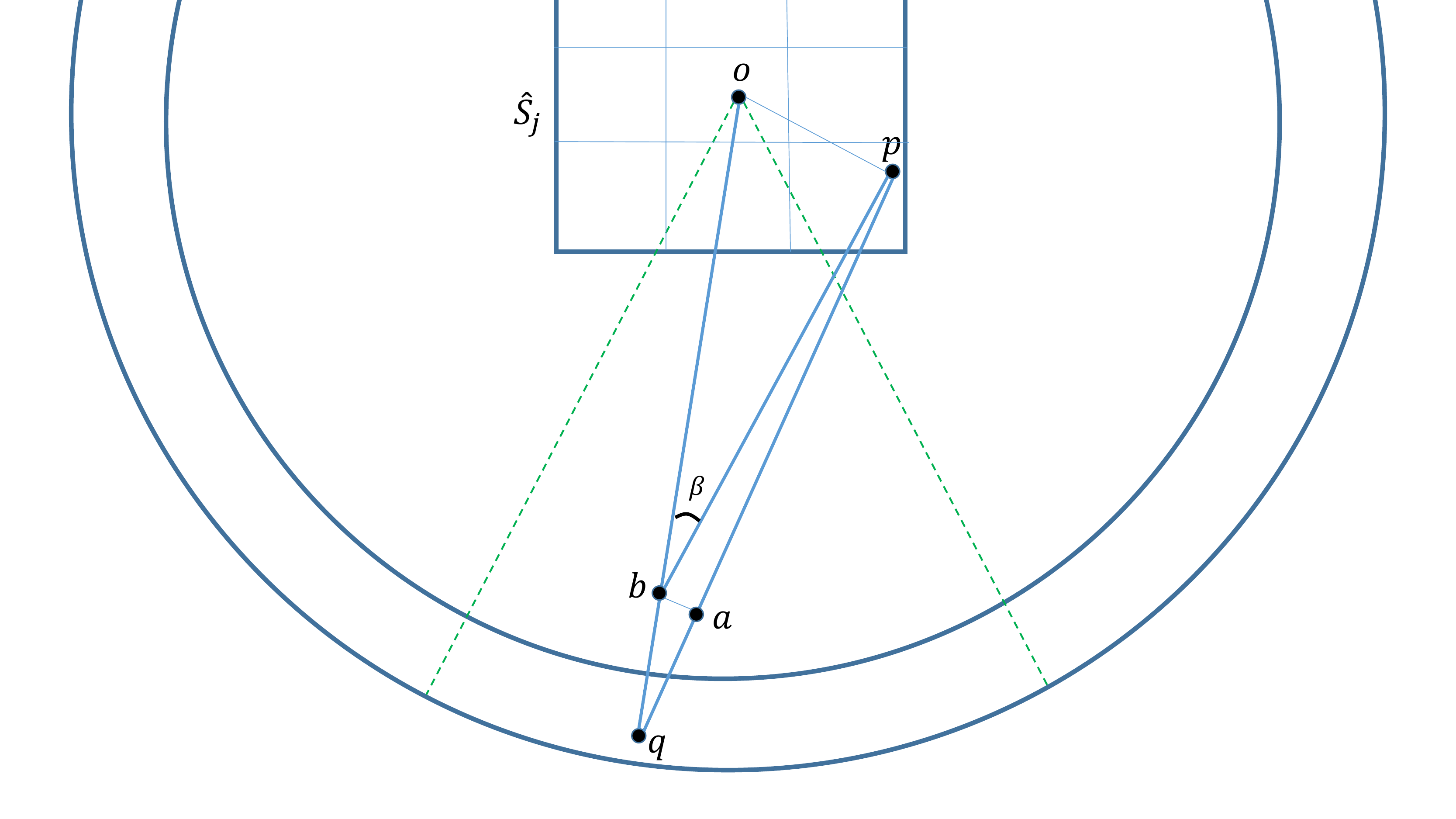}
  \end{center}
  \caption{\small An illustration of the proof of Lemma \ref{lem:circle1}. }
  \label{fig:circles1}
\end{figure}

For part (b), let $o$ be the center of $S_j$, and let $b$ be the point on the ray through $oq$ such that
$|pb|=r$ (for the assumed ranges of $r$ and $\eps$, $b$ is unique). Assume that $b$ lies between $o$ and $q$;
the case where $b$ lies beyond $q$ is handled analogously.
It suffices to show that $|qb| \le 2\eps$, which is the $\rho$-range of a small polar rectangle $R'$.
See Figure \ref{fig:circles1}.

Let $\beta = \angle obp$. Applying the Law of Sines in the triangle $\triangle obp$, we get that
$$
\frac{|op|}{\sin\beta} = \frac{|pb|}{\sin\angle pob} \ge r \ .
$$
Hence $\sin\beta \le \frac{|op|}{r} \le \frac{3\delta_1/\sqrt{2}}{r}$.

Since we assume that $\delta_1\ll r$, we may also assume, as in the proof of Lemma~\ref{lem:circle2},
that $\beta < \frac{3\delta_1}{r} \ll 1$. Hence $\angle pbq > \pi/2$, and thus
$|pq| > |pb|=r$. Let $a$ be the point on $pq$ for which $|pa|=r$.
Applying the Law of Sines in the triangle $\triangle abq$, we get that
\begin{equation} \label{eqn-qba}
\frac{|qb|}{\sin\angle qab} = \frac{|qa|}{\sin\angle qba} .
\end{equation}
By assumption, $|qa| \le\eps$. Also, $\angle qba = \pi - \beta - \angle pba$.
In the isosceles triangle $\Delta pab$, we have
$$
\frac{\pi}{2} > \angle pba = \frac{\pi}{2} - \frac{\angle bpa}{2} >
\frac{\pi}{2} - \frac{\beta}{2} ,
$$
and therefore
$$
\frac{\pi}{2} - \beta < \angle qba < \frac{\pi}{2} - \frac{\beta}{2} .
$$
Substituting these bounds in Equation (\ref{eqn-qba}) we get
$$
|qb| = \frac{|qa|\sin\angle qab}{\sin\angle qba} < \frac{\eps}{\cos(\pi/2-\angle qba)} < \frac{\eps}{\cos\beta} ,
$$
which is smaller than $2\eps$ when $\beta$ is small enough, that is, when $r$ is sufficiently larger than $\delta_1$.
$\Box$

\medskip
\noindent
{\bf The algorithm.}
The preceding analysis yields the following straightforward implementation,
analogous to the one of Section \ref{sec:pl2d}. We first compute, for each point $p\in P$,
the square $S_i$ that contains it and we find, for each circle $c_q\in C$, the squares that it crosses.
This gives us all the sets $P_i$, $C_i$.
We then iterate over the small squares in the partition of $S$. For each such
square $S_i$, we construct the dual partitioning (in polar coordinates)  of the resulting dual rectangle
$R_i$ into the smaller rectangles $R'$.
As above, we find, for each dual point $q$, for which $c_q\in C_i$, the small rectangle $R'$ that
contains it, and, for each dual circle $c_p$, for $p\in P_i$, the small rectangles that it crosses.
We now report, for each small rectangle $R'$, all the pairs $(p,q)$ for which
$q$ lies in $R'$ and $c_p$ crosses either $R'$ or one of the small rectangles
lying directly above and below $R'$ (in the $\rho$-direction, if they exist).
We repeat this over all small primal squares $S_i$ and all respective small rectangles
$R'$. Note that a pair $(p,q)$ may be reported more than once in this procedure,
but its multiplicity is at most some small absolute constant.

As in the case of lines, the running time of this algorithm is $O(n/\delta_1 + m/\delta_2 + k)$,
where $k$ is the output size. By Lemmas~\ref{lem:circle2} and \ref{lem:circle1}, every pair
$(p,q)$ at distance in $[r-\eps,r+\eps]$ will be reported, and every reported pair lies at distance
in $[r-\alpha\eps,r+\alpha\eps]$, for some small absolute constant $\alpha$,
provided that we enforce the constraints $\delta_1\ll r$, $\delta_2\le\sqrt{\eps/r}$, and
$\delta_1\delta_2 = \sqrt{2}\eps$.

As in Section \ref{sec:pl2d}, to minimize the running time, while satisfying
$\delta_1\delta_2 = \sqrt{2}\eps$, we want to pick
$$
\delta_1 = \sqrt{\frac{2n\eps}{m}} , \quad\quad\text{and} \quad\quad
\delta_2 = \sqrt{\frac{m\eps}{n}} .
$$
The other two constraints amount to requiring that $\eps n \ll mr^2$ and $mr\le n$. That is,
\begin{equation} \label{constraints}
\frac{\eps}{r^2} \ll \frac{m}{n} \le \frac{1}{r} .
\end{equation}
Since the problem is symmetric in $P$ and $Q$, we may assume that $m\le n$
(otherwise we simply flip the roles of $P$ and $Q$). Hence the right inequality in (\ref{constraints}) holds (recall that we assume that $r \leq 1/2$).
If the other inequality does not hold, say, $m/n \le 100\eps/r^2$, we skip the primal stage, apply only the dual
partitioning, with $\delta_2 = \Theta(\eps)$, and get the bound $O\left(\frac{m}{\eps} + n + k\right) = O\left(\frac{n}{r^2}+k \right)=O(n+k)$.
(Recall that we assume that $r$ is bounded from below by a constant $\upsilon$.)

In the remaining case, (\ref{constraints}) holds, and then $\delta_1$, $\delta_2$
are both $\le 1$, as is easily checked, and then the bound is $O(\sqrt{mn}/\sqrt{\eps} + k)$.
Including the symmetric case $m\ge n$, we get the following theorem, which improves upon
Theorem \ref{th:circles1-2d} when the values of $m$ and $n$ are ``unbalanced''.

\begin{theorem} \label{th:circles2-2d}
Let $P$ and $Q$ be two sets of $m$ and $n$ points, respectively, in the unit disk $B$, and let
 $\upsilon \le r \le 1/2$ for a constant $\upsilon$. We can report all pairs $(p,q)\in P\times Q$ for which $||pq|-r|\le\eps$,
in
$$
O\left(m + n + \frac{\sqrt{mn}}{\sqrt{\eps}} + k \right) ,
$$
time, where $k$ is the actual number of pairs that we report; all pairs at distance in $[r-\eps,r+\eps]$
will be reported, and every reported pair lies at distance in
$[r-\alpha\eps,r+\alpha\eps]$, for some absolute constant $\alpha$.
\end{theorem}

\section{Near-neighbor point-circle configurations} \label{sec:arb_circ2d}

In this section we study the near-neighbor problem for points and arbitrary circles, extending
the results from the previous section. Specifically, we are given a set $P$ of $m$ points in
the unit disk $B$ in the plane, and a set $C$ of $n$ circles intersecting $B$, where we assume
that the radii of the circles in $C$ all lie in a fixed interval $[r_1,r_2]$, for $\eps \le r_1 \le 1/2\le r_2$.
We want to
compute the $\eps$-approximate incidences between $P$ and $C$.

We solve this problem by a two-stage partition scheme, similar to those used before, except that one
stage takes place in the plane, and the other in three dimensions.

The first stage is more or less identical to that used in Section~\ref{sec:pc2dII}.
That is, we assume that $P$ is contained in the unit square $S=[0,1]^2$.
We fix two real positive parameters $\delta_1$, $\delta_2$, whose values will be set later.
We partition $S$ into $\frac{1}{\delta_1^2}$ pairwise openly disjoint smaller
squares, each of side length $\delta_1$.
We enumerate these squares as $S_1,\ldots,S_{1/\delta_1^2}$, and let $\hat{S}_i$ denote
the union of $S_i$ and the (at most) eight squares adjacent to $S_i$.
Let $P_i$ denote the set of all points of $P$ that lie in $\hat{S}_i$, and let $C_i$
denote the set of all the circles $c\in C$ that cross $S_i$.
Put $m_i:=|P_i|$ and $n_i:=|C_i|$, for $i=1,\ldots,1/\delta_1^2$. We have
$\sum_i m_i = O(m)$, and $\sum_i n_i = O(n/\delta_1)$.

The second stage is different, because the varying values of $r$ do not allow us to apply the
simple duality that we used in Section~\ref{sec:pc2dII}. Instead we first move to a different
notion of distance between a point and a circle, which is the \emph{power}
(see, e.g.,~\cite{Coxeter}).
The power of a point $p$ with respect to a circle $c$ centered at $q$ with radius $r$ is
$\Pi(p,c) = |pq|^2-r^2$.

The notions of Euclidean distance and power are closely related:
Let $p$ be a point and $c$ a circle centered at a point $q$ with radius $r$.
Notice that  $\dd(p,c) = | |pq|-r |$ and
$$
|\Pi(p,c) | = | |pq|^2-r^2 | = \dd(p,c)(|pq|+r) =\dd(p,c)(2r+\dd(p,c)) .
$$
Hence, we always have
\begin{equation} \label{eqn-lowr1}
 |\Pi(p,c)| \ge 2r\dd(p,c) \ge 2r_1\dd(p,c) ,
\end{equation}
 and if
 $\dd(p,c) \le r$, which certainly holds for all circles $c$ which are approximately incident to $p$,
we have
\begin{equation} \label{eqn-upr2}
|\Pi(p,c)| \le 3r \dd(p,c) \le 3r_2 \dd(p,c).
\end{equation}
By Equation (\ref{eqn-upr2}), for every  pair $p, c$ such that $\dd(p,c)\le \eps$
we have that $|\Pi(p,c)| \le 3r_2 \eps$. On the other hand, for a pair
$p, c$ such that $|\Pi(p,c)| \le 3r_2 \eps$ we know, by Equation (\ref{eqn-lowr1}), that
$\dd(p,c)\le 3r_2 \eps/2r_1$. Thus, our task now is
to report all pairs $p,c$ such that $|\Pi(p,c)| \le 3r_2 \eps$.

We use the standard lifting transform where each point $p=(x,y)\in\reals^2$ is mapped to the
point $(x,y,x^2+y^2)$ on the paraboloid $z=x^2+y^2$ in 3-space, and each circle $c$
with center $q=(q_1,q_2)$ and radius $r$ is mapped to the plane $z=2q_1x+2q_2y + (r^2-q_1^2- q_2^2)$. The vertical distance between the images of $p$ and $c$  is
$$
\big| x^2+y^2 - 2q_1x-2q_2y + q_1^2 + q_2^2-r^2 \big| =
\big| |pq|^2 - r^2 \big| = |\Pi(p,c)| .
$$
We now dualize 3-space by mapping points to planes and planes to points, in a standard manner that
preserves vertical distances between points and planes. We get a set $P_i^*$ of $m_i$
planes and a set $C_i^*$ of $n_i$ points, and want to report all point-plane pairs
at vertical distance $\le 3r_2\eps$. This is handled exactly as in the second stage
in Section~\ref{sec:pp3d}.

Specifically, since each circle $c\in C_i$ crosses $S_i$, its distance from the center $o_i$
of $S_i$ is at most $\delta_1/\sqrt{2}$, so the vertical distance between the plane $o_i^*$
and the point $c^*$ is at most $3r_2\cdot \delta_1/\sqrt{2} < 3r_2\delta_1$.
Moreover, the $xy$-projection of $c^*$ is $2q$, which lies in a suitable annulus
centered at $2o_i$ with radii proportional to $r_1$ and $r_2$; this holds if we
require that $\delta_1 < r_1$, say. For simplicity, enclose this annulus by an axis-parallel
square $R_0$ of side length $cr_2$, for a suitable constant $c$,
and let $R$ denote the parallelepiped bounded between the two planes that are shifts of $o_i^*$
by $\pm 3r_2\delta_1$ and having $R_0$ as its $xy$-projection.

We now partition $R$ into $O(1/\delta_2^3)$ small homothetic copies, each scaled down by $\delta_2$.
Each small region $R'$ has an $xy$-projection of size $cr_2\delta_2\times cr_2\delta_2$,
and its vertical width (in the $z$-direction) is $3r_2\delta_1\delta_2$.

For each small region $R'$, we report all the pairs $(p,c)\in P_i\times C_i$ for which
$c^*\in R'$ and $p^*$ crosses either $R'$ or one of the two regions above and below $R'$
with the same $xy$-projection. Each dual plane $p^*$ crosses $O(1/\delta_2^2)$ small regions.

Applying the arguments used in the case of points and planes in $\reals^3$, given
in Section~\ref{sec:pp3d}, and in particular ensuring that $\delta_1\delta_2 = \eps$, we conclude that the algorithm correctly reports all pairs $(p,c)$
for which $\dd(p,c) \le\eps$, and that each pair that it reports satisfies
$\dd(p,c) \le\alpha\eps$, for a suitable constant $\alpha$. The overall running time is
$$
O\left( \frac{n}{\delta_1}+\frac{m}{\delta_2^2} + k \right) ,
$$
where $k$ is the number of pairs that we report. To optimize this bound,
we choose $\delta_1$ and $\delta_2$ to satisfy
$$
\frac{n}{\delta_1} = \frac{m}{\delta_2^2} , \quad\quad\text{and}\quad\quad \delta_1\delta_2 = \eps ,
$$
that is,
$$
\delta_1 = \left(\frac{\eps^2n}{m}\right)^{1/3} , \quad\quad\text{and}\quad\quad
\delta_2 = \left(\frac{\eps m}{n}\right)^{1/3} .
$$
We require that $\delta_1 < r_1$, $\delta_2 \le 1$. In case $\delta_1>r_1$, that is, $m < n\eps^2r_1^3$,
we skip the first stage, and run the second stage over the full data, with $\delta_2 = \eps$, resulting
in running time $O\left( \frac{m}{\eps^2} + n + k \right) = O(n+k)$.
Similarly, in case $\delta_2>1$, that is, $n < m\eps$, we perform only the first stage,
with $\delta_1 = \eps$, and the running time is then
$O\left( m + \frac{n}{\eps} + k \right) = O(m+k)$.
Otherwise the running time is $O\left(m^{1/3}n^{2/3}/\eps^{2/3} + k\right)$.

Thus, we have obtained the following theorem.
\begin{theorem} \label{th:circles-2d}
Let $P$ be a set of $m$ points in the unit disk $B$ in $\reals^2$,
let $C$ be a set of $n$ circles of radii in the range $[r_1,r_2]$, for some positive constants $\eps \le r_1\le 1/2 \le r_2$,
that cross $B$, when $\eps>0$ is a prescribed error parameter.
We can report all pairs $(p,c)\in P\times C$ for which $\dd(p,c)\le\eps$ in time
$$
O\left(m+n+ \frac{m^{1/3}n^{2/3}}{\eps^{2/3}} + k \right) ,
$$
where $k$ is the actual number of pairs that we report; all pairs at distance at most $\eps$
will be reported, and every reported pair lies at distance at most $\alpha\eps$ for some constant $\alpha$ (proportional to $r_2/r_1$).
\end{theorem}

\section{Reporting all nearly congruent pairs in three dimensions} \label{sec:cong_spheres3d}

In this section we consider the three-dimensional version of the problem studied in
Section~\ref{sec:circ2d}. That is, we are given  sets $P$ and $Q$ of $m$ and  $n$ points, respectively,
 in the unit ball $B$
in $\reals^3$, and parameters $0< \upsilon \le r \le 1/2$, for a constant $\upsilon$,  and wish to report all pairs
$(p,q)\in P^2$ such that $\dd(p,q)\in [r-\eps,r+\eps]$. As usual, we allow more pairs to be
reported, but require that each pair $(p,q)$ that we report satisfies
$\dd(p,q)\in [r-\alpha\eps,r+\alpha\eps]$, for some absolute constant $\alpha$.
This is the approximate incidences problem between $P$ and spheres of radius $r$ centered at the points of $Q$,

As in Section \ref{sec:circ2d}, there are two alternative solutions, one using the technique of Indyk et al.~\cite{Indyk:1999},
and one using duality. In the following we derive Theorem \ref{th:balls1-3d} using
Indyk et al.'s approach. We omit the tedious derivation using duality which would give a result analogous to
Theorem \ref{th:circles2-2d} (with $\eps$ rather than $\sqrt{\eps}$ in the denominator).

Let $\sigma_o$ denote the sphere of radius $r$ centered at the origin $o$.
We can cover $\sigma_o$ with $O(1/\eps)$ congruent caps, each of opening
angle $\sqrt{\eps}$,
so that
no point of $\sigma_o$ is covered by more than $O(1)$ caps.
Let  $U$ be the set of directions from $o$ to the centers of these caps\footnote{One can do this by packing  disjoint caps of opening angle $\sqrt{\eps}/2$ on $\sigma_o$, and taking $U$ as the set of directions to the centers of these caps.}, $|U| = O(1/\eps)$.
In the following we fix one direction $u\in U$, which, without loss of generality, we assume
to be the positive $z$-direction.

Let $\frownacc{u}$ denote the cap of $\sigma_o$ with $u$ as a central direction.
Let $A_{\frownacc{u}}$ be a cap portion of a spherical shell centered at $o$,
with inner radius $r-\eps$ and outer radius $r+\eps$, which is the intersection of the entire such shell
with the cone with apex $o$, axis $u$, and opening angle $\sqrt{\eps}$.
Let $R_{\frownacc{u}}$ denote the smallest enclosing axis-parallel box of $A_{\frownacc{u}}$
(Figure~\ref{fig:circle3} can serve as a schematic two-dimensional illustration of this setup).

Let $R_{\frownacc{u}}(q)$ be $R_{\frownacc{u}}$ translated by a vector (point) $q\in Q$.
Let $\R$ denote the collection of the boxes $R_{\frownacc{u}}(q)$, for $q\in Q$.
Note that the members of $\R$ are translates of one another.

We now construct a grid $G$ whose cells are translates of $R_{\frownacc{u}}$, assign each point
of $P$ to the cell of $G$ containing it,
and assign each point $q\in Q$ to the at most (exactly, in general position) eight cells that $R_{\frownacc{u}}(q)$
overlaps. We then report, over all grid cells, all the pairs $(p,q)\in P\times Q$ that are assigned to the same cell.

We repeat this procedure for each of the $O(1/\eps)$ orientations in $U$, and the overall output of the algorithm
is the union of the outputs for the individual orientations. The overall running time is $O((m+n)/\eps + k)$,
where $k$ is the number of distinct pairs that we report.
The term $O(k)$ is justified because each pair $(p,q)$
is reported once for each shell-cap of $q$ such that the box
$R^*_{\frownacc{u}}$ which is a  homothetic copy $R_{\frownacc{u}}$ scaled by a factor of $3$
about its center, contains $p$. It is easy to check that if $p$ lies in $R^*_{\frownacc{u}}$ then
the angle between $\vec{qp}$ and $u$ is $O(\sqrt{\eps})$ so there could be only $O(1)$ such directions $u$.

As in Section \ref{sec:pc2dI}, the correctness of this algorithm is a consequence of the following lemma.

\begin{lemma} \label{lem:main-ballR3}
(a) Let $q$ be a point at distance $\le\eps$ from $\sigma_o$, so that the point of $\sigma_o$ nearest
to $q$ lies in ${\frownacc{u}}$. Then $q\in R_{\frownacc{u}}$.

\smallskip
\noindent
(b) Let $R^*_{\frownacc{u}}$ denote the homothetic copy $R_{\frownacc{u}}$ scaled by a factor of $3$
about its center. Then every point in $R^*_{\frownacc{u}}$ is at distance $\le \alpha\eps$ from $\sigma_o$, for
a suitable small absolute constant $\alpha$.
\end{lemma}
\noindent{\bf Proof.}
(a) is trivial since in this case $q$ must lie in $A_{\frownacc{u}}$ and therefore also in  $R_{\frownacc{u}}$.

We establish (b) by giving a lower (resp., upper) bound on the shortest (resp., longest) distance of a point in
$R^*_{\frownacc{u}}$ from $o$.

Clearly, the point of $R_{\frownacc{u}}$ closest to $o$ is the center point $\mu$ of its bottom face.
This point $\mu$ lies on the  cross section of $R_{\frownacc{u}}$ with the $yz$-plane.
This cross section is congruent to the rectangle $R_{\frownacc{ab}}$ of Figure \ref{fig:circle3} bounding the annulus
$A_{\frownacc{ab}}$ around an arc $\frownacc{ab}$ with opening angle $\sqrt{\eps}$.

Arguing as in the proof of Lemma \ref{lem:main-distR2}, the distance of $\mu$ from $o$ is at least
$r-2\eps$, and therefore the distance of the center $\mu^*$ of the bottom face of $R^*_{\frownacc{u}}$
from $o$ is at least $r-5\eps$.

The points of $R^*_{\frownacc{u}}$ farthest from $o$ are the four vertices of its top face $f$.
Arguing as in the proof of Lemma~\ref{lem:main-distR2}, the center point of $f$ lies at distance
at most $r+4\eps$ from $o$. The side length of $f$ is the same as the side length of its
cross section with the $yz$-plane, which is at most $3\sqrt{\eps}$, as in the proof of
Lemma \ref{lem:main-distR2}, so the distance of a vertex of $f$ from its center point
is at most $3\sqrt{\eps/2}$. By the Pythagorean theorem, we obtain that the distance of $o$
from a vertex of $f$ is at most
$$
\left( (r+4\eps)^2 + 9\eps/2 \right)^{1/2} =
\left( r^2 + (8r+9/2)\eps + 16\eps^2 \right)^{1/2} < r + \alpha\eps ,
$$
for a suitable constant $\alpha$ that depends on $r$ (analogously to the analysis in Section~\ref{sec:pc2dI}).
$\Box$

\smallskip
The following theorem summarizes the main result of this section.
\begin{theorem} \label{th:balls1-3d}
Let $P$ and $Q$ be two sets, of respective sizes $m$ and $n$, in the unit ball $B$ in $\reals^3$,
and let $0<\upsilon \le r \le 1/2$  for some constant $\upsilon$.
For any small $\eps$, we can report all pairs $(p,q)\in P\times Q$ for which $\dd(p,q)\in [r-\eps,r+\eps]$, in
$$
O\left( \frac{m+n}{\eps} + k \right)
$$
time, where $k$ is the actual number of distinct pairs that we report. All pairs at distance
in $[r-\eps,r+\eps]$ will be reported, and every reported pair lies at distance
in $[r-\alpha\eps,r+\alpha\eps]$, for some constant $\alpha$ that depends on $r$.
\end{theorem}

\noindent{\bf Remark.}
Note that both techniques work in any dimension, more or less verbatim. Consider for example Indyk et al.'s technique. One major difference is that
the size of the set $U$ of directions in $d$ dimensions is $O(1/\eps^{(d-1)/2})$, so the algorithm runs in time
$O((m+n)/\eps^{(d-1)/2} + k)$; the naive grid-based approach, discussed in the introduction, would take $O(m+n/\eps^{d-1} + k)$.
There is also the issue of applying the Pythagorean theorem, where the factor $\sqrt{2}$ has to be
replaced by $\sqrt{d-1}$. The rest of the analysis goes more or less unchanged.

\section{Reporting all point-line neighbors in three dimensions} \label{sec:pl3d}

Let $P$ be a set of $m$ points in the unit ball $B$ in three dimensions,
let $L$ be a set of $n$ lines that cross $B$, and let $\eps>0$ be a given error parameter.
In this section we present an algorithm for the approximate incidence reporting problem
involving $P$ and $L$.

We represent each line in $\reals^3$ by the pair of equations\footnote{%
  We assume without loss of generality that no line is orthogonal to the $x$-axis.}
$y=ax+b$, $z=cx+d$.
Let $\ell$ be the line $y=ax+b$, $z=cx+d$, and let $p=(\xi,\eta,\zeta)$ be a point in $\reals^3$.
We approximate $\dd(p,\ell)$ by slicing space by the plane $\pi_p:\;x=\xi$, and by computing the
distance between the points $p$ and $\ell_p:= \ell\cap\pi_p = (\xi,a\xi+b,c\xi+d)$.

As in Section \ref{sec:pl2d}, for this approximation to be good, the angle between $\ell$ and
the $x$-direction should not be too large. To ensure this, similarly to what we did in
Section~\ref{sec:pp3d}, we partition $L$ into $O(1)$ subfamilies,
such that, for each subfamily $L'$ there exists a direction $u'$ such that the angle between
$u'$ and each line of $L'$ is at most $\pi/4$. We apply the construction to each subset
$L'$ separately (with respect to all the points in $P$). To apply it for a specific $L'$,
we first rotate $\reals^3$ so that $u'$ becomes the $x$-direction. In what follows,
we fix one subfamily, continue to denote it as $L$, and assume that $u'$ is indeed the $x$-direction.

We make the following two easy observations (compare with the analysis in Section~\ref{sec:pp3d}).
\begin{lemma} \label{lem:ac}
The slopes $a$ and $c$ of any line $y=ax+b$, $z=cx+d$ in $L$ satisfy $a^2+c^2\le 1$.
\end{lemma}
\noindent{\bf Proof.}
The parametric representation of the line $y=ax+b$, $z=cx+d$ is
$\{ (0,b,d)+t(1,a,c) \mid t\in\reals\}$. By our assumption, the angle $\gamma$
between the vectors $(1,a,c)$ and $(1,0,0)$ is at most $\pi/4$. Hence we have
$$
\cos\gamma = \frac{1}{\sqrt{1+a^2+c^2}} \ge {\rm cos}(\pi/4) = \frac{1}{\sqrt{2}} \ ,
$$
from which the lemma follows.
$\Box$

\begin{figure}[htb]
\begin{center}
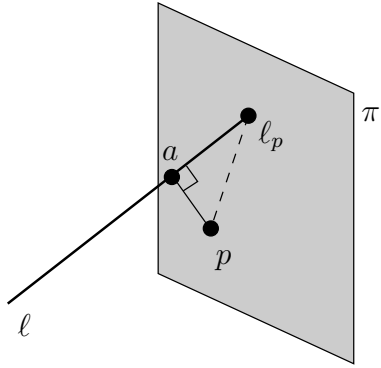
\caption{We approximate the distance between $p$ and $\ell$ by the distance between $p$ and $\ell_p$. }
\label{fig:ellpi}
\end{center}
\end{figure}

\begin{lemma} \label{lem:approx}
Let $p=(\xi,\eta,\zeta)$ be a point in $P$, let $\ell$ be a line in $L$, and let $\ell_p:= \ell\cap\pi_p$ where $\pi_p$ is the plane $x=\xi$. Then
$\dd(p,\ell_p) \le \sqrt{2}\dd(p,\ell)$.
\end{lemma}
\noindent{\bf Proof.}
Let $a$ be the point on $\ell$ closest to $p$, and consider the triangle $\triangle a\ell_p p$.
Since the angle between $\ell$ and the $x$-direction is at most $\pi/4$, the angle $\theta_0$
between $\ell$ and its projection on $\pi_p$ is at least $\pi/4$.
Since $\theta_0$ is the smallest angle between $\ell$ and any line in $\pi_p$,
it follows that the angle $\angle a\ell_p p$ is also at least $\pi/4$, and therefore
$$
\frac{\dd(p,\ell)}{ \dd(p,\ell_p)} = \frac{\dd(p,a)}{ \dd(p,\ell_p)}
\ge \sin(\pi/4) = \frac{1}{\sqrt{2}}\ ,
$$
as claimed. See Figure \ref{fig:ellpi}.
$\Box$

As in the preceding sections, we replace $B$ by the unit cube $S=[0,1]^3$ and we assume $P \subseteq S$.
Then we apply the following two-stage partitioning procedure.

\medskip
\noindent
{\bf The primal stage.}
We fix two parameters $\delta_1$, $\delta_2$, whose values will be set later.
we first partition $S$ into ${1}/{\delta_1^3}$ pairwise openly disjoint smaller cubes,
each of side length $\delta_1$. Enumerate these cubes as $S_1,S_2,\ldots,S_{1/\delta_1^3}$.
For $i=1,\ldots,1/\delta_1^3$, let $P_i$ denote the set of all points of $P$ that lie in $S_i$
or in one of the (at most) eight cubes that surround $S_i$ and have the same $x$-projection as $S_i$.
Let $L_i$ denote the set of all the lines of $L$ that cross $S_i$.
Put $m_i:=|P_i|$ and $n_i:=|L_i|$, for $i=1,\ldots,1/\delta_1^3$. We have $\sum_i m_i \le 9m$,
and $\sum_i n_i \le 3n/\delta_1$, as is easily checked (to cross from a cube to an adjacent cube,
the line has to cross one of the $3/\delta_1$ planes that define the grid).

\medskip
\noindent
{\bf The dual stage.}
For each such small cube $S_i$, we now pass to a parametric dual four-dimensional space
(with coordinates $(x,y,z,w)$), in which we represent each line $\ell\in L_i$,
given by $y=ax+b$, $z=cx+d$, by the point $\ell^* = (a,b,c,d)$, and represent
each point $p = (\xi,\eta,\zeta)\in P_i$ by the 2-plane (in $\reals^4$)
$$
p^* = \{ (a,b,c,d) \mid a\xi+b = \eta,\; c\xi+d = \zeta \} ;
$$
$p^*$ is the locus of all points dual to lines that pass through $p$.

We define the distance in the dual space between a point $\ell^* = (a,b,c,d)$
and a plane $p^*$, for a primal point $p=(\xi,\eta,\zeta)$,
to be the distance between $\ell^*$ and the point $(a,\eta-a\xi,c,\zeta-c\xi)$,
which is the intersection of $p^*$ with the plane defined by $x=a$ and $z=c$.
In the primal space, the point $(a,\eta-a\xi,c,\zeta-c\xi)$ corresponds to a line parallel
to $\ell$ that passes through $p$.  The following lemma is immediate from this definition.
\begin{lemma} \label{lem:dist}
The distance between $\ell^*$ and $p^*$, as defined above, is equal to $\ddv(p,\ell_p)$ in the primal space.
\end{lemma}

Fix a small cube $S_i$, and assume without loss of generality that $S_i=[0,\delta_1]^3$.
Let $\ell$ be a line in $L_i$, given by $y=ax+b$, $z=cx+d$.
Since we assume that the angle between
each each line of $L$ and the $x$-axis is at most $\pi/4$,
the $y$- and $z$-spans of the intersection of $\ell$ with the slab $0 \le x \le \delta_1$
are each at most $\delta_1$. This implies that $-\delta_1 \le b \le 2\delta_1$, and
$-\delta_1 \le d \le 2\delta_1$.

It also follows from Lemma \ref{lem:ac} that $|a|,|c|\le 1$. Therefore, in the dual
parametric four-dimensional space, $\ell^*$ lies in the box
$R$ given by
\begin{align*}
-1 & \le a,\;c \le 1 \\
-\delta_1 & \le b,\;d \le 2\delta_1 .
\end{align*}
We now partition $R$ into $1/\delta_2^4$ smaller boxes, each of which is a homothetic copy of $R$
scaled down by $\delta_2$. Concretely, each smaller box $R'$ is
congruent to the box $[0,2\delta_2]\times [0,3\delta_1\delta_2]\times [0,2\delta_2]\times [0,3\delta_1\delta_2]$.

\begin{lemma} \label{lem:close1}
For each small box $R'$, if $\ell^* = (a_\ell,b_\ell,c_\ell,d_\ell)$ is a dual point
(of some $\ell \in L_i$) in $R'$ and $p^*$ is a dual plane (of some point $p=(\xi,\eta,\zeta)\in P_i$)
that crosses $R'$ or one of its (at most eight) surrounding boxes of the same $xz$-range,
then $\dd(p,\ell) \le 8\sqrt{2}\delta_1\delta_2$.
\end{lemma}
\noindent{\bf Proof.}
Assume without loss of generality that $S_i$ is the cube $[0,\delta_1]^3$ and that $R'$ is
the box
$[0,2\delta_2]\times [0,3\delta_1\delta_2]\times [0,2\delta_2]\times [0,3\delta_1\delta_2]$.
Since $\ell^*$ is in $R'$ we have
\begin{align}
0 & \le a_\ell,\; c_\ell \le  2 \delta_2 \label{eq:222a} \\
0 & \le b_\ell,\; d_\ell \le 3\delta_1\delta_2 . \label{eq:222b}
\end{align}
Let $(a,b,c,d)$ be a point in $p^*\cap \hat{R}'$, where $\hat{R}'$ is $R'$
or one of its surrounding boxes of the same $xz$-range. By definition of $p^*$, we have
$b=\eta-a\xi$ and $d=\zeta-c\xi$. Since $(a,b,c,d)\in \hat{R}'$, we have\footnote{%
  Note that we include here adjacent regions that lie outside $R$, because suitable shifts
  of them would arise when $R'$ is a generic small region.}
\begin{align}
0 & \le a,\; c \le 2 \delta_2 \label{eq:333a} \\
-3\delta_1\delta_2 & \le \eta-a\xi,\; \zeta-c\xi  \le 6\delta_1\delta_2 . \label{eq:333b}
\end{align}
Finally, since $p\in P_i$, we have
\begin{align}
0 & \le \xi \le \delta_1 \label{eq:555} \\
-\delta_1 & \le \eta \le 2\delta_1 \nonumber \\
-\delta_1  & \le \zeta \le 2\delta_1 .  \nonumber
\end{align}
We have
$$
\dd(p,\ell) \le \dd(p,\ell_p) = \left( (a_\ell\xi + b_\ell - \eta)^2 + (c_\ell\xi + d_\ell - \zeta)^2 \right)^{1/2} .
$$
Let us estimate the first term
$a_\ell\xi + b_\ell - \eta$ in the square root; the second term is estimated in a fully analogous manner.
We have
$$
a_\ell\xi + b_\ell - \eta = (a_\ell-a)\xi+ b_\ell + a\xi- \eta .
$$
By Equations (\ref{eq:222a}) and (\ref{eq:333a}), $(a_\ell-a) \in [-2\delta_2,2\delta_2]$,
and by Equation (\ref{eq:555}), $\xi \in [0,\delta_1]$, so $(a_\ell-a)\xi \in [-2\delta_1\delta_2,2\delta_1\delta_2]$.
By Equation (\ref{eq:222b}), $b_\ell\in [0,3\delta_1\delta_2]$, and by equation (\ref{eq:333b}),
$a\xi- \eta \in [-6\delta_1\delta_2,3\delta_1\delta_2]$. Adding up these estimates, we get
$$
a_\ell\xi + b_\ell - \eta \in [-8\delta_1\delta_2,8\delta_1\delta_2] ,
$$
and, by a fully symmetric argument,
$$
c_\ell\xi + d_\ell - \zeta \in [-8\delta_1\delta_2,8\delta_1\delta_2] .
$$
Hence $\dd(p,\ell) \le 8\sqrt{2}\delta_1\delta_2$, as asserted.
$\Box$

\medskip
For the following lemma, we constrain $\delta_1$ and $\delta_2$ to satisfy $\delta_1\delta_2=\eps$,
and $\delta_2 \le 1/\sqrt{2}$.
\begin{lemma} \label{reptok}
(a) Let
$p=(\xi,\eta,\zeta)$ be a point in $P$ and $\ell$ be a line in $L$,
given by $y=a x + b$, $z=c x + d$, such that $\dd(p,\ell) \le \eps$.
Let $S_i$ be the small primal cube containing $p$. Then $\ell$ must cross either $S_i$ or one
of the at most eight cubes that surround $S_i$ and have the same $x$-projection as $S_i$.
Therefore, there exists a $j$ such that $(p,\ell)\in P_j\times L_j$.

\smallskip
\noindent
(b) Let $p$ and $\ell$ be as in (a), let $j$ be such that $(p,\ell)\in P_j\times L_j$, and let $R'$ be the dual small
region in $\reals^4$ (that arises in the dual processing of $S_j$) that contains $\ell^*$.
Then the dual plane $p^*$ must cross either $R'$ or an adjacent small region with the same $xz$-projection.
\end{lemma}
\noindent{\bf Proof.}
(a) The line $\ell$ crosses the plane $x=\xi$ at the point
$q = (\xi, a\xi+b, c\xi+d)$ which, by Lemma \ref{lem:approx}, lies at
distance at most $\sqrt{2}\eps$ from $p$. That is, $q$ lies in a cube with the same
$x$-projection as $S_i$, at distance at most $\sqrt{2}\eps = \sqrt{2}\delta_1\delta_2$
from $S_i$, so, for  $\delta_2\le 1/\sqrt{2}$, it must lie either in $S_i$ or
in one of the eight surrounding cubes with the same $x$-projection, as claimed.

\smallskip
\noindent
(b) Assume without loss of generality that
$R'=[0,2\delta_2]\times [0,3\delta_1\delta_2]\times [0,2\delta_2]\times [0,3\delta_1\delta_2]$.
The unique intersection point of $p^*$ with the plane $\pi:\; x=a,\;z=c$ is the point
$\lambda^* = (a, \eta - a\xi, c, \zeta - c\xi)$. Within $\pi$,
the absolute value of the $y$-shift (resp., $w$-shift) between $\ell^*$ and $\lambda^*$ is
$\big| a\xi + b - \eta \big|$ (resp., $\big| c\xi + d - \zeta \big|$).
By construction and by Lemma \ref{lem:approx}, each of these quantities is at most
$d(p,\ell_p) \le \sqrt{2}\eps = \sqrt{2}\delta_1\delta_2$. This implies
that the region $R''$ containing $\lambda^*$ must be adjacent to $R'$ and of the same $xz$-projection as $R'$.
$\Box$

\medskip
\noindent
{\bf The algorithm.}
The preceding analysis yields the following straightforward implementation. We compute the sets
$P_i$, $L_i$, for $i=1,\ldots,1/\delta_1^3$, in overall $O(m+n/\delta_1)$ time. Then, for each small
cube $S_i$, we consider the partitioning of the resulting dual box $R$
into the smaller boxes $R'$. As above, we find, for each $\ell\in L_i$, the small region that contains
the dual point $\ell^*$, and, for each $p\in P_i$, the small regions that the dual plane $p^*$ crosses.
This takes $O(n_i+m_i/\delta_2^2)$ time.

We now report, for each small region $R'$, all the pairs $(p,\ell)\in P_i\times L_i$ for which
$\ell^*$ lies in $R'$ and $p^*$ crosses either $R'$ or one of the at most eight small regions
that surround $R'$ and have the same $xz$-range. We repeat this over all small
cubes $S_i$ and all respective small regions $R'$. The running time of the algorithm is
$$
O\left( m + \frac{n}{\delta_1} + \sum_{i=1}^{1/\delta_1^3} \left(n_i + \frac{m_i}{\delta_2^2}\right) + k \right) =
O\left(\frac{n}{\delta_1} +  \frac{m}{\delta_2^2} + k \right) ,
$$
where $k$ is the number of distinct point-line pairs that we report. By Lemmas~\ref{lem:close1} and \ref{reptok},
every pair $(p,\ell)$ at distance at most $\eps=\delta_1\delta_2$ will be reported, and every reported
pair lies at distance at most $8\sqrt{2}\eps$. Moreover, no pair is reported more than a constant number of times.
To minimize the running time overhead, as a function of $\eps$, we choose $\delta_1$ and $\delta_2$ to satisfy
$$
\frac{n}{\delta_1} = \frac{m}{\delta_2^2} , \quad\quad\text{and}\quad\quad \delta_1\delta_2 = \eps ,
$$
that is,
$$
\delta_1 = \left(\frac{\eps^2n}{m}\right)^{1/3} \quad\quad\text{and}\quad\quad
\delta_2 = \left(\frac{\eps m}{n}\right)^{1/3} .
$$
As in all the preceding cases, we need to require that $\delta_1$ and $\delta_2$ are both at most $1$,
and in fact we want $\delta_2$ to be at most $1/\sqrt{2}$.
If one of these conditions does not hold, we skip the corresponding primal or dual stage,
set the other parameter to $\eps$, and conclude, as is easily verified, that the running time is $O(m+n+k)$.
Otherwise, the running time is $O\left(m^{1/3}n^{2/3}/\eps^{2/3} +k \right)$.

In summary, we obtain the following result.
\begin{theorem} \label{th:lines-3d}
Let $P$ be a set of $m$ points in the unit ball $B$ in $\reals^3$, let $L$ be a set of
$n$ lines in $\reals^3$ that cross $B$, and let $\eps>0$ be a prescribed error parameter.
We can report all pairs $(p,\ell)\in P\times L$ for which $\dd(p,\ell)\le\eps$ in
$$
O\left(m+n+ \frac{m^{1/3}n^{2/3}}{\eps^{2/3}} +k \right)
$$
time, where $k$ is the actual number of distinct pairs that we report. All pairs at distance
$\le\eps$ will be reported, and every reported pair lies at distance at most $8\sqrt{2}\eps$.
\end{theorem}


\section{Reporting all point-circle neighbors in three dimensions} \label{sec:pc3d}

In preparation for our final algorithm, of finding all nearly congruent copies of a given triangle
in a set of $n$ points in $\reals^3$, we first solve the following problem. Let $P$ be a set of
$m$ points in the unit ball $B$ in $\reals^3$, let $C$ be a set of $n$ congruent circles in $\reals^3$
of radius $r \le 1/2$ that cross $B$, and let $\eps \ll r$ be a prescribed error parameter.
We present an efficient algorithm for the approximate incidence reporting problem for $P$ and $C$.

We relax the problem further, as follows. For each circle $c\in C$, denote by $\lambda_c$ the \emph{axis}
of $c$, which is the line that passes through the center of $c$ and is orthogonal to the plane $\pi_c$ containing $c$.
We partition $C$ into $O(1)$ subsets, corresponding to some canonical set $U$ of $O(1)$ directions
in $\reals^3$, so that we associate with each $u\in U$ all the circles $c\in C$ for which the angle
between $u$ and $\lambda_c$ is at most some small but constant value $\theta_0$; in general, this is
not a partition, but we turn it into a partition by assigning each circle $c$ to an arbitrary single
set from among those it belongs to. We apply the following procedure separately for each of these subsets,
and focus on a single such set, where we assume, without loss of generality, that the corresponding
direction $u$ is the $z$-vertical direction. For simplicity of notation, continue to denote the
corresponding subset of $C$ as $C$.

Now fix a circle $c\in C$, and denote by $T_c$ the torus that is the Minkowski sum of $c$
and the ball $B_\eps$ of radius $\eps$ centered at the origin; our goal is to report all
points in $P\cap T_c$, for each circle $c\in C$.

Let $c_0$ be the circle of radius $r$ in the $xy$-plane centered at the origin. We partition
$T_{c_0}$ into sectors by roughly $\pi/\sqrt{\eps}$ planes through the $z$-axis such that the dihedral angle between each pair
of consecutive planes is $\sqrt{\eps}$. We enumerate the sectors as $S_{c_0}^{(1)},\ldots,S_{c_0}^{(\kappa)}$,
where $\kappa = \pi/\sqrt{\eps}$.
We now focus on one such sector $S_{c_0}^{(j)}$ and to simplify the notation drop the index $j$ from
$S_{c_0}^{(j)}$ hereafter.
Let $\gamma_{c_0}$ be the arc $c_0\cap S_{c_0}$ of $c_0$, and let
$s_{c_0}$ be the chord connecting the endpoints of $\gamma_{c_0}$.
We rotate $\reals^3$ around the $z$-axis so that $s_{c_0}$ is parallel to the $x$-axis.
Let $Q_{c_0}$ be the smallest cylinder enclosing $S_{c_0}$ whose axis is parallel to
$s_{c_0}$. The cross section of $Q_{c_0}$ with the $xy$-plane is a rectangle $R_{c_0}$,
similar to the one shown in Figure \ref{fig:circle3} (where $\overset{\frown}{ab}$ is now $\gamma_{c_0}$).
As the calculations in Section \ref{sec:pc2dI} show, the width of $R_{c_0}$ is $< 3\eps$ and its length
is $<\sqrt{\eps}$ (and these bounds are tight up to an absolute constant factor).
In other words, the radius of $Q_{c_0}$ is at most $\eps'=1.5\eps$.

Now consider a circle $c\in C$. Let $\pi$ be the translation of the $xy$-plane
that maps the origin to the center of $c$. Tilt $\pi$ by some angle $\theta$, which is at
most $\theta_0$, around its intersection line with $\pi_c$ until it coincides with $\pi_c$;
this also makes the image of $c_0$ coincide with $c$, and that of $T_{c_0}$ coincide with $T_c$.
Let $S_c$ be the sector of $T_c$ that is the corresponding image of $S_{c_0}$.
Let $\gamma_c$, $s_c$, $Q_c$ be the corresponding arc, segment, and cylinder, respectively.

Our approximation goal now is to report all pairs $(p,c)$, such that $p$ lies in $Q_c$,
and do so for every $c\in C$ and for every sector $S_{c_0}^{(j)}$ of $S_{c_0}$. By construction, every pair $(p,c)$ satisfying $\dd(p,c)\le\eps$
(such that $c$ is in our current subset) is such that $p$ lies in at least one of the
$O(1/\sqrt{\eps})$ cylinders $Q_c$ of $c$, and will therefore be reported.
We perform this task for every sector index $j$.
(As is easily checked, $p$ lies in at most two cylinders $Q_c$, for the same circle $c$, so
$(p,c)$ will be reported at most twice.)

This new problem is reminiscent of the problem of reporting all near point-line pairs in three dimensions,
as presented in Section~\ref{sec:pl3d}, with a major difference that instead of lines we have segments
(namely, the bounded axes of the cylinders $Q_c$). Our next step further reduces our problem to the
point-line scenario.

\begin{figure}[htbp]
  \begin{center}
    \includegraphics[scale=0.75]{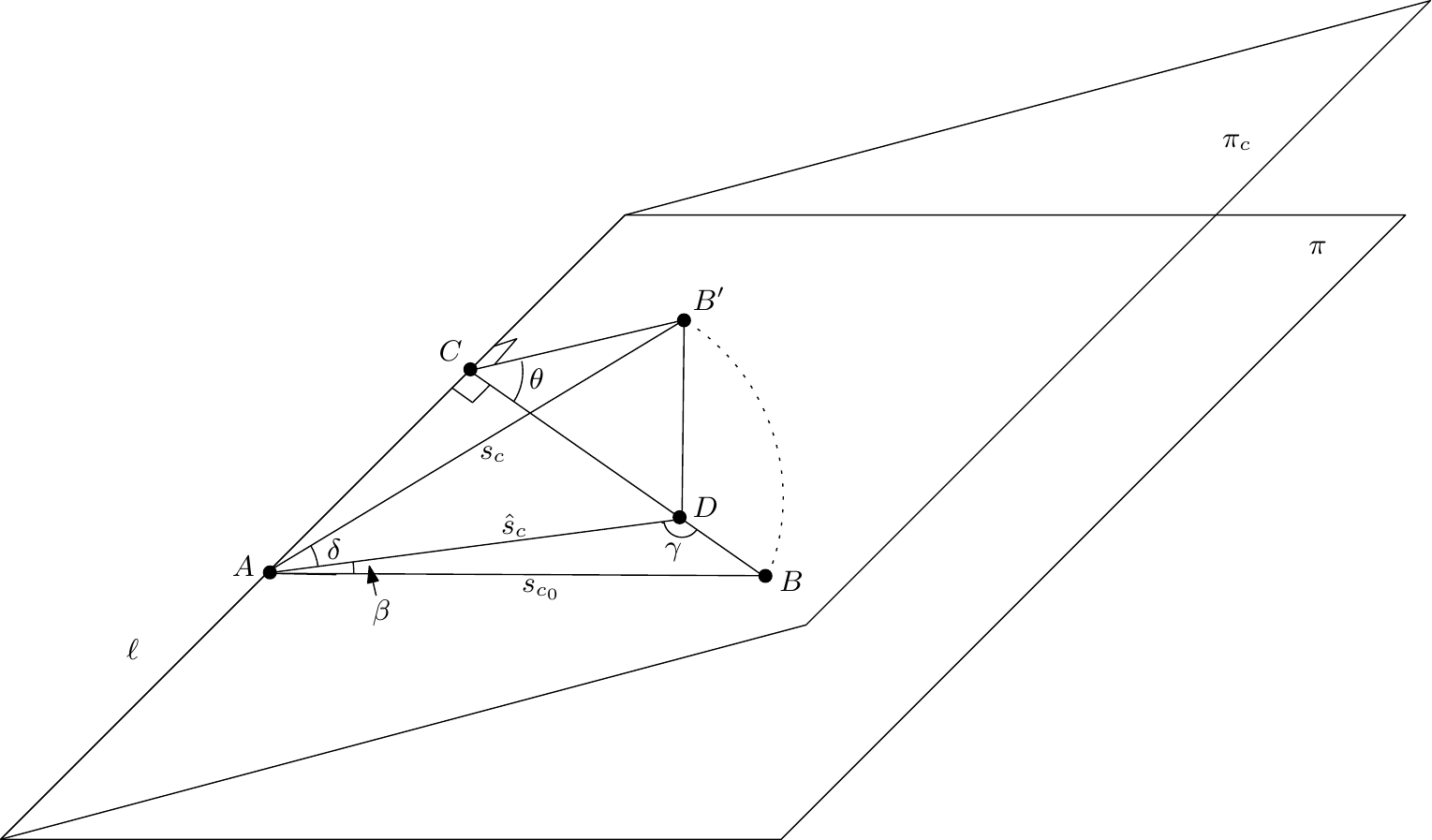}
  \caption{The proof that $\hat{s}_c$ is close to $s_{c_0}$.}
  \label{fig:stero1}
  \end{center}
\end{figure}

Let $\hat{s}_c$ denote the projection of $s_c$ onto the $xy$-plane. We claim that
$|s_{c_0}|\cos \theta_0\le |\hat{s}_c| \le |s_{c_0}|$, and that the angle $\beta$ between
$\hat{s}_c$ and $s_{c_0}$ is small (satisfying $\beta = O(\theta_0^2)$).
Indeed (refer to Figure \ref{fig:stero1}), assume without loss of generality that
$s_{c_0}=AB$ is such that its endpoint $A$ lies on $\ell=\pi\cap\pi_c$. The rotation of $\pi$
by angle $\theta\le\theta_0$ around $\ell$ brings $B$ to a point $B'\in\pi_c$, and $s_{c_0}$ to $s_c=AB'$.
Let $D$ be the projection of $B'$ on $\pi$, and let $C$ be the point on $\ell$ nearest to $D$.
The projection $\hat{s}_c$ of $s_c$ onto $\pi$  is $AD$, and
the projection of $CB'$ onto $\pi$ is $CD$.  As is easily verified, $D$ lies on the segment $CB$,
$\angle B'CD=\theta$, and $|CB'|=|CB|$.
Since $|AB'|=|AB|$, we have $|\hat{s}_c| \le |s_{c_0}|$. On the other hand, $|AB'|\ge|CB'|$,
so $\delta = \angle B'AD \le \theta$. Hence, $|\hat{s}_c| = |AB'|\cos\delta \ge |s_{c_0}|\cos\theta_0$.
To estimate $\beta$, we have $|BD|= |BC| - |B'C|\cos\theta = |BC|(1-\cos\theta) = |BC|\cdot O(\theta_0^2)$.
Hence, $\sin\beta = |BD|\sin(\angle ADB)/|AB| \le |BD|/|CB| = O(\theta_0^2)$.

By our assumption that $s_{c_0}$ is parallel to the $x$-axis, we obtain that each of
the projections $\hat{s}_c$, for $c\in C$, is of length at most $r\sqrt{\eps}<\sqrt{\eps}$,
and is almost parallel to the $x$-axis, forming with it an angle which is $O(\theta_0^2)$.

We partition $\reals^3$ into vertical slabs that are orthogonal to the $x$-axis, and are
of width equal to $|s_{c_0}| \le \sqrt{\eps}$. Our bounds on the lengths of $\hat{s}_c$ and their angles with
$s_{c_0}$ imply that the axis $a_c$ of each cylinder $Q_c$ crosses only
$O(1)$ slabs. Furthermore, if
we stretch $a_c$ by a factor of $3$ about its center then the projection of this larger segment onto
the $x$-direction completely covers each of the slabs that $a_c$ intersects, assuming that $\theta_0$
is sufficiently small.

\begin{figure}[htb]
\begin{center}
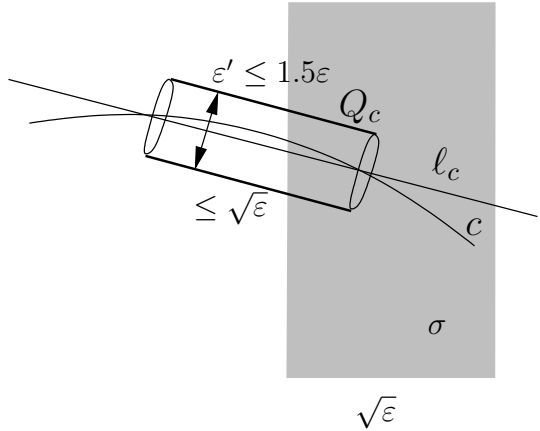
\caption{For each slab $\sigma$ that $Q_c$ intersects we include the line $\ell_c$ through the
axis of $Q_c$ in the point-line approximate incidence reporting problem associated with $\sigma$. }
\label{fig:inslab}
\end{center}
\end{figure}

We fix a slab $\sigma$, take the subset of the cylinders that cross $\sigma$,
and replace each such cylinder $Q_c$ by the (entire) line $\ell_c$ containing its axis.
See Figure \ref{fig:inslab}.
Then we apply the algorithm in Section~\ref{sec:pl3d} to the set of these lines and to the
set of points of $P$ within $\sigma$, with an error parameter $\eps' \le 1.5\eps$,
equal to the common radius of all the cylinders $Q_c$.
Let $Q^*_c$ be a cylinder obtained from $Q_c$ by stretching its axis about its center by a factor of $3$
and increasing its radius to $8\sqrt{2} \eps$ (see Theorem \ref{th:lines-3d}).
By the discussion above, the output will contain all pairs $(p,\ell_c)$ such that $p\in Q_c$,
and every output pair will satisfy $p\in Q^*_c$.

As is easily checked, each pair $(p,c)$ can be reported once for each sector $j$, as
$p$ is contained in only one slab. The same pair can be reported at most twice, in the
subproblems associated with a pair of adjacent sectors, when $p$ is contained in both
corresponding (and slightly overlapping) cylinders $Q^*_c$.

We apply the algorithm in Section~\ref{sec:pl3d} to every slab that contains at least one point and is
crossed by at least one cylinder. For each slab $\sigma$, let $m_\sigma$ denote the
number of points of $P$ in $\sigma$, and let $n_\sigma$ denote the number of cylinders that
cross $\sigma$. As noted, we have $\sum_\sigma m_\sigma \le m$, and $\sum_\sigma n_\sigma = O(n)$.
The time required by the algorithm in Section~\ref{sec:pl3d}, for a slab $\sigma$, is
$$
O\left( m_\sigma + n_\sigma + m_\sigma^{1/3}n_\sigma^{2/3}/\eps^{2/3} + k_\sigma^{(u,j)} \right) ,
$$
where $k_\sigma^{(u,j)}$ is the number of reported pairs for the subproblem associated with $u$, $j$, and $\sigma$.
Summing over all slabs $\sigma$ (with $u,j$ still fixed), and using H\"older's inequality, the total running time is
$$
O\left( m + n + m^{1/3}n^{2/3}/\eps^{2/3} + k^{(u,j)} \right) ,
$$
where $k^{(u,j)}$ is the overall number of reported pairs for the subproblem associated with $u$ and $j$.
Summing over the $O(1)$ values of $u$, and the $O(1/\sqrt{\eps})$ sector indices $j$, and observing
(as already noted) that a pair $(p,c)$ is reported at most $O(1)$ times, we get a total running time of
$$
O\left( \frac{m + n}{\eps^{1/2}} + \frac{m^{1/3}n^{2/3}}{\eps^{7/6}} + k \right) ,
$$
where $k$ is the number of (distinct) reported pairs, over all possible choices of all parameters.

\medskip
\noindent
{\bf Correctness.}
Similar to the previous cases, the correctness is established in the following lemma.
\begin{lemma} \label{lem:circ3d-1}
(a) Each pair $(p,c) \in P\times C$ satisfying $\dd(p,c)\le\eps$ will be reported by the algorithm.

\smallskip
\noindent
(b) Each pair $(p,c) \in P\times C$ reported by the algorithm satisfies $\dd(p,c)\le\alpha\eps$,
for a suitable absolute constant $\alpha$.
\end{lemma}
\noindent{\bf Proof.}
(a) Let $(p,c) \in P\times C$ be such that $\dd(p,c)\le\eps$. Let $u\in U$ be the direction associated with $c$
and consider the sectors and slabs associated with $u$.

Let $S_c$ be the sector of $T_c$ that contains $p$ (there exists at least one,
and in general exactly one such sector). Clearly, the enclosing cylinder $Q_c$ also contains $p$.
Let $\sigma$ be the slab that contains $p$ in the subproblem associated with the sector $S_c$.
Then $Q_c$ too must cross $\sigma$, and the
correctness of the algorithm in Section~\ref{sec:pl3d} implies that $(p,c)$ will be reported.

(b) Let $(p,c) \in P\times C$ be a pair that we report, at some subproblem with the direction $u\in U$,
associated with $c$, at some the sector $S_c$, and at the corresponding slab $\sigma$ that contains $p$.
Refer to Figure~\ref{fig:inslab}. By the discussion above, we have $p\in Q^*_c$.
The proof is completed by arguing that any point $p \in Q^*_c$ is at distance at most
$\alpha \eps$ from $c$ for some absolute constant $\alpha$. This can be done by estimating the distance
of the furthest point in $Q^*_c$ from the center of $c$ using the
Pythagorean theorem as in the proofs of Lemmas \ref{lem:main-distR2}
and \ref{lem:main-ballR3}.
$\Box$

\smallskip
In summary, we obtain the following result.
\begin{theorem} \label{ptcirc3d-I}
Let $P$ be a set of $m$ points in the unit ball $B$ of $\reals^3$,
let $C$ be a set of $n$ congruent circles in $\reals^3$, of common radius $r \le 1/2$,
that cross $B$, and let $\eps\ll r$ be a prescribed error parameter.
We can report all pairs $(p,c)\in P\times C$ that satisfy $\dd(p,c)\le\eps$, in time
$$
O\left( \frac{m + n}{\eps^{1/2}} + \frac{m^{1/3}n^{2/3}}{\eps^{7/6}} + k \right) ,
$$
where $k$ is the number of (distinct) pairs that we report. Each pair satisfying
$\dd(p,c)\le\eps$ will be reported, and each reported pair satisfies
$\dd(p,c)\le\alpha\eps$, for some absolute constant $\alpha$.
Moreover, each pair is reported at most $O(1)$ times.
\end{theorem}

\section{Reporting all nearly congruent triangles} \label{sec:tri}

In this section we put to work the algorithms in Sections \ref{sec:cong_spheres3d} and \ref{sec:pc3d} (see also (e) and (g) in Section \ref{sec:intro}), to obtain an efficient solution of the first
step in solving the approximate point pattern matching problem in $\reals^3$ (see its review in the introduction), where we are given
a sampled ``reference'' triangle $\Delta abc$, for a triple of points $a$, $b$, $c$ in the first set $A$,
and a prescribed error parameter $\eps>0$. Our goal is to report all triples $p,q,o$ in the second set $B$
that span a triangle ``nearly congruent'' to $\Delta$; that is, triples that satisfy
\begin{equation} \label{nearc}
\big| |pq|-|ab| \big| \le \eps, \quad
\big| |po|-|ac| \big| \le \eps, \quad\text{and}\quad
\big| |qo|-|bc| \big| \le \eps .
\end{equation}
We require that all such triples are reported, but we also allow to report triples that satisfy (\ref{nearc}) with $\alpha \eps$ on the right-hand sides rather than
$\eps$, for some fixed absolute constant $\alpha$. Let $ab$ be the longest edge of $\Delta$.
We require that $\beta\le |ab| \le 1/2$ for some fixed constant $\beta$. We also require
that the height $h$ of $\Delta$ from $c$ (perpendicular to $ab$) is larger than some fixed constant $s$.
We assume that $\beta, s \gg \eps$. Our approximation guarantee $\alpha$ increases as $\beta$ and $s$ decrease.

We first report all pairs $(p,q) \in B^2$ such that $\big| |pq|-|ab| \big| \le \eps$, using the algorithm in Section \ref{sec:cong_spheres3d} which involves incidences between congruent spheres and points).
This takes $O(n/\eps + N)$ time, where $N$ is the number of pairs that we report.
Let $\Pi$ denote the set of reported pairs. We know that all the desired pairs are included in $\Pi$,
and that every pair $(p,q)$ in $\Pi$ satisfies $\big| |pq|-|ab| \big| \le \alpha'\eps$, for some
absolute constant $\alpha'$. We prune $\Pi$, leaving in it only pairs $(p,q)$
satisfying $\big| |pq|-|ab| \big| \le \eps$. We continue to denote the resulting set as $\Pi$,
and its size by $N$.

\begin{figure}[!h]
  \centering
   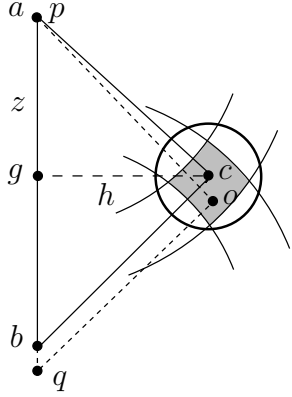
    \caption{The reference triangle $\Delta abc$ aligned with $\Delta pqo$. The shaded region is $K$.
    The surrounding disk is a cross section of the torus $T_{p,q}$. \label{fig:retorus}}
\end{figure}

Let $(p,q)$ be a pair in $\Pi$. Any point $o$ that satisfies $\big| |po|-|ac| \big| \le \eps$ and
$\big| |qo|-|bc| \big| \le \eps$
lies in the intersection $K=K_{p,q}$ of two spherical shells, one centered at $p$ with radii $|ac|\pm\eps$,
and one centered at $q$ with radii $|bc|\pm\eps$.
The following lemma allows us to replace $K$ by a torus that is congruent to a fixed torus that depends only on $\Delta$.
See Figure \ref{fig:retorus}.

\begin{lemma} \label{lem:torus}
Assume that $\Delta$ is sufficiently fat, in the sense that $\beta\le |ab| \le 1/2$ and $h\ge s$,
for some absolute positive constants $\beta$, $s$ that satisfy $\eps\ll \beta,s$.
Then there exists a circle $\gamma_{p,q}$ of radius $h$ such that
$K$ is contained in the torus $T_{p,q}$ that is the Minkowski sum of $\gamma_{p,q}$ and a
ball of radius $\eps' \le \delta\eps$ around the origin, where the constant $\delta$
depends on $\beta$ and $s$.
\end{lemma}

\noindent{\bf Proof.}
Denote the lengths of the edges of the  triangle $\Delta abc$ by $u=|ab|$, $v=|ac|$ and $w=|bc|$.
Let $g$ the point where $h$ meets $ab$ and let $z=|ag|$.
We have $z^2 + h^2 = v^2$ and $(u-z)^2 + h^2 = w^2$, from which we obtain that
$z = \frac{ u^2 + v^2 - w^2 }{2u}$, and we denote this expression as $z=z(u,v,w)$.
Consider an alignment of $\Delta$ within the plane of $\Delta pqo$, such that $a$ coincides with $p$ and $ab$
overlaps $pq$. Let $g$ now be a point on $pq$ at distance $z$ from $p=a$. Then $c$ lies on the
circle $\gamma_{p,q}$ of radius $h$, centered at $g$, and contained in the plane perpendicular to
$pq$ through $g$. See Figure~\ref{fig:retorus}.

Fix some point $o\in K$. We claim that $o$ must be at distance $\le \delta \eps$ from $\gamma_{p,q}$,
for some fixed constant $\delta$ that depends on $\beta$ and $s$. Indeed, since $(p,q)\in\Pi$ and
$o\in K$, we can write
$|pq| = u + \eps_1$, $|po| = v + \eps_2$, and $|qo| = w + \eps_3$,
where $|\eps_i|\le\eps$ for $i=1,2,3$.

Consider the alignment of $\Delta$ with $\Delta pqo$, as above, and imagine that we
perturb the edges $ab$, $ac$, and $bc$ of $\Delta$ by $\eps_1$, $\eps_2$, and $\eps_3$, respectively,
so that $\Delta$ is continuously deformed into $\Delta pqo$.
We claim that $o$ cannot move too far  as a result of this deformation so the distance between $o$ and $c$ must be small.

To see this, let $h'$ be the height of $\Delta pqo$ from $o$, let $g'$ be the point at which
$h'$ meets $pq$, and let $z' = |pg'|$. We claim that $|z' - z| \le \delta \eps$ and $|h'-h| \le \delta \eps$
for some absolute constant $\delta$. To see this, using the function $z=z(u,v,w)$ defined above, we have
$z'=z(u+\eps_1,v+\eps_2,w+\eps_3)$, and routine calculations show that, for $\eps$ sufficiently small, we have
$|z' - z| = O(|\nabla z(u,v,w)\cdot (\eps_1,\eps_2,\eps_3)|) \le \delta' \eps$, where $\delta'$ depends on $\beta$.

Similarly, by Heron's formula, we can think of $h$ as a function $h(u,v,w)$, given by
$$
h(u,v,w) = \frac{2{\rm Area}(\Delta)}{u} = \frac{2\sqrt{\tau(\tau-u)(\tau-v)(\tau-w)}}{u} \ ,
$$
where $\tau=\frac12 (u+v+w)$. Then $h'=h(u+\eps_1,v+\eps_2,w+\eps_3)$, and, by another routine calculation,
$|h' - h| = O(|\nabla h(u,v,w)\cdot (\eps_1,\eps_2,\eps_3)|) \le \delta'' \eps$, for another constant
$\delta''$ that depends on $\beta$ and $s$. (Simple calculations show that $|\nabla h|$ becomes smaller
as $s$ increases.) Take $\delta = \sqrt{(\delta')^2+(\delta'')^2}$, and the lemma follows.
$\Box$

\smallskip
We have thus reached the following scenario. We have a set $\T$ of $N$ congruent tori $T_{p,q}$,
for $(p,q)\in \Pi$, and a set $B$ (the original one) of $n$ points.
By construction, each triple $(p,q,o)$ that defines a triangle for which (\ref{nearc}) holds, satisfies $o\in T_{p,q}$.
Using our algorithm for point-circle near neighbors in $\reals^3$, as reviewed in Section \ref{sec:pc3d},
we can report all the triples $(p,q,o)$ such that $o\in T_{p,q}$,
in time $O\left( {n + N}/{\eps^{1/2}} + {n^{1/3}N^{2/3}}/{\eps^{7/6}} + k \right)$,
where $k$ is the number of (distinct) triples that we report; each of the desired triples is reported,
and each triple that we report is such that the distance from $o$ to $\gamma_{p,q}$ is at most $\alpha\eps$ for some
other fixed constant $\alpha > \delta$. Therefore each triple which we report satisfies (\ref{nearc}) with
$\alpha \eps$ on the right-hand sides, rather than $\eps$.
In summary, we have:
\begin{theorem} \label{th:main}
Let $B$ be a set of $n$ points in the unit ball in $\reals^3$. Let $\Delta abc$ be
a fixed reference triangle and let $\eps$ an error parameter, so that $\Delta$ and $\eps$ satisfy
the constraints specified in Lemma \ref{lem:torus}.
We can then report all triples $(p,q,o)\in B^3$ that span a triangle nearly
congruent to $\Delta$, in the sense of (\ref{nearc}), in time
$
{\displaystyle
\left( {n + N}/{\eps^{1/2}} + {n^{1/3}N^{2/3}}/{\eps^{7/6}} + k \right) ,}
$
where $N$ is the number of pairs reported by our algorithm for approximate congruent pairs in $\reals^3$
(presented in Section \ref{sec:cong_spheres3d}),
applied to $P$ with distance $|ab|$, the largest edge length of $\Delta$, and $k$ is the number of
(distinct) triples that the algorithm in this section reports; each of the desired triples is reported,
and each triple that we report satisfies (\ref{nearc}) with $\alpha\eps$ replacing $\eps$, where
$\alpha$ is a suitable absolute constant. Each pair is reported at most $O(1)$ times.
\end{theorem}

\section{Implementation and experiments}
\label{sec:exp}

To test the effectiveness of the methodology proposed in this paper, we  implemented the algorithm
of Section \ref{sec:pl2d}, for incidences of points and lines in the plane, and tested it on real and random data.
We compared its performance to three other approaches
that are used in practice, and were mentioned in the introduction. Specifically we compared the algorithms:

\medskip
\noindent{{\sf Naive}: Based on constructing a grid of cell size $\eps$ only in the primal plane.
Its running time is $O(m+n/\eps+k)$.}

\medskip
\noindent{{\sf Naive-duality}: Use the naive approach when $m > n$. Otherwise apply the naive solution in the dual plane.
The running time is $O(m+n+min(m,n)/\eps+k)$.}

\medskip
\noindent{{\sf Large-$n$ (the dense case)}: The alternative solution of Aiger and Kedem~\cite{AK10}.
Its running time is $O\left(m+n+\frac{1}{\eps^2}\log\frac{1}{\eps}+k\right)$.}

\medskip
\noindent{{\sf Efficient-duality} ({\sf Efficient} for short in the plots): This is our solution, with running time $O(m+n+\sqrt{nm}/\sqrt{\eps}+k)$.}

The output size $k$, which appears
in the four time bounds listed above, is not a fixed quantity, because it depends on the specific algorithm
being used. More precisely, for a fixed input instance, denote by $k_{true}$ the real output size, which is
the number of pairs at distance at most $\eps$ apart. Each algorithm encounters its own superset of these pairs,
and its running time degrades linearly with the size of this superset.

 As a matter of fact, our {\sf Efficient-duality}
algorithm tends to have a larger value of $k$, because each of its primal and dual steps makes some worst-case
assumptions that affect the size of the grid cells that are used, allowing more pairs to be reported.
The {\sf Efficient-duality} algorithm
might report pairs at distance up to $5\sqrt{2}\eps$ (see Section \ref{sec:pl2d}),
whereas each pair reported by the {\sf Naive} implementation is only at distance at most
$2\sqrt{2}\eps$, as is easily checked.

Our random data set consisted of
$n$  points drawn uniformly at random in the unit square and $n$ random lines crossing that square, for various values of
$n$. For this data
the value of $k$ tends to increase quadratically in the respective factor $5\sqrt{2}\eps$, $2\sqrt{2}\eps$,
and the difference could become significant when $\eps$ is large.

\begin{figure}[htb]
  \centering
  \subcaptionbox{}{\includegraphics[width=0.45\textwidth]{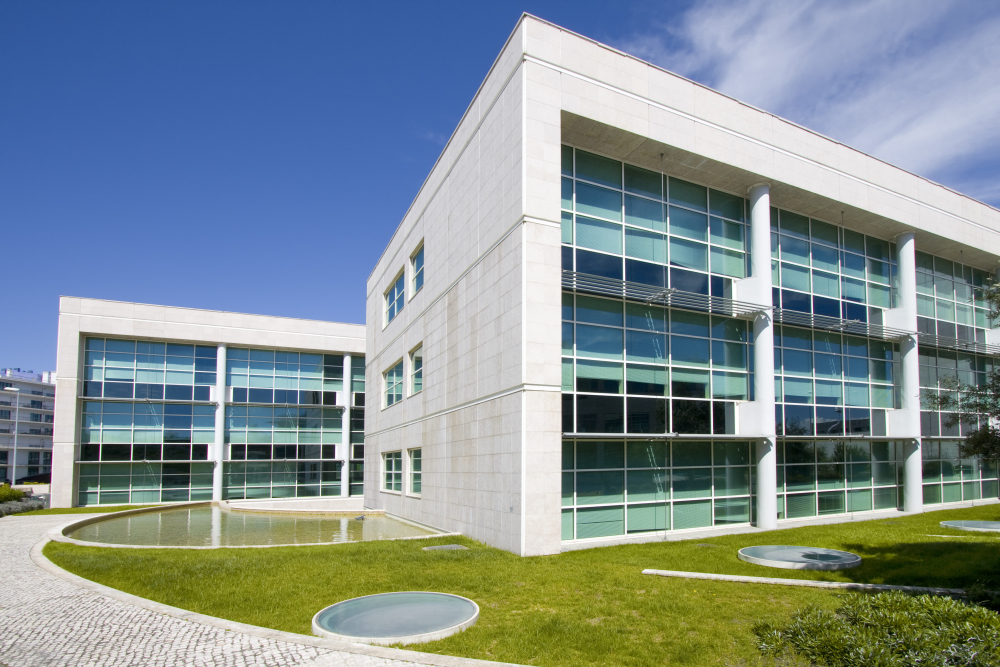}}
  \subcaptionbox{}{\includegraphics[width=0.45\textwidth]{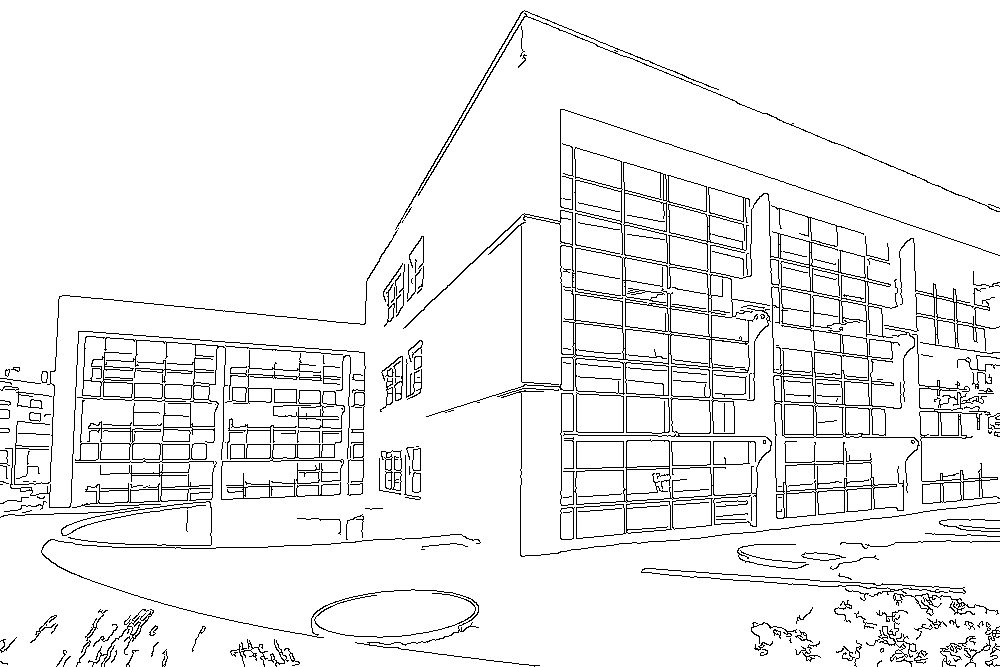}}
  \caption{The input with real data: (a) The image. (b) The detected edges, from which we sample our input points.}
  \label{fig:images}
\end{figure}

Our real data set was
 extracted from the image depicted in
Figure~\ref{fig:images}(a). That is, we have applied a standard edge detection procedure to this image,
resulting in the edges depicted in Figure~\ref{fig:images}(b), from which we have sampled our points.
The lines that we use were obtained by sampling pairs of these points, in the hope that some of the sampled
lines will be very close to the actual edges, and will be detected as such by the approximate incidence reporting algorithms.
In other words, the experiments that we have conducted on this data were made with the application of robust model
fitting in mind; see later in this section.

In real data, if we use an algorithm that allows pairs at distance up to $\alpha\eps$ to be reported,
we expect that the number $k$ of reported (i.e., inspected) pairs will grow only linearly in $\alpha$.

Our results are as follows.

\begin{figure}[!htb]
  \centering
  \subcaptionbox{}{\includegraphics[width=0.45\textwidth]{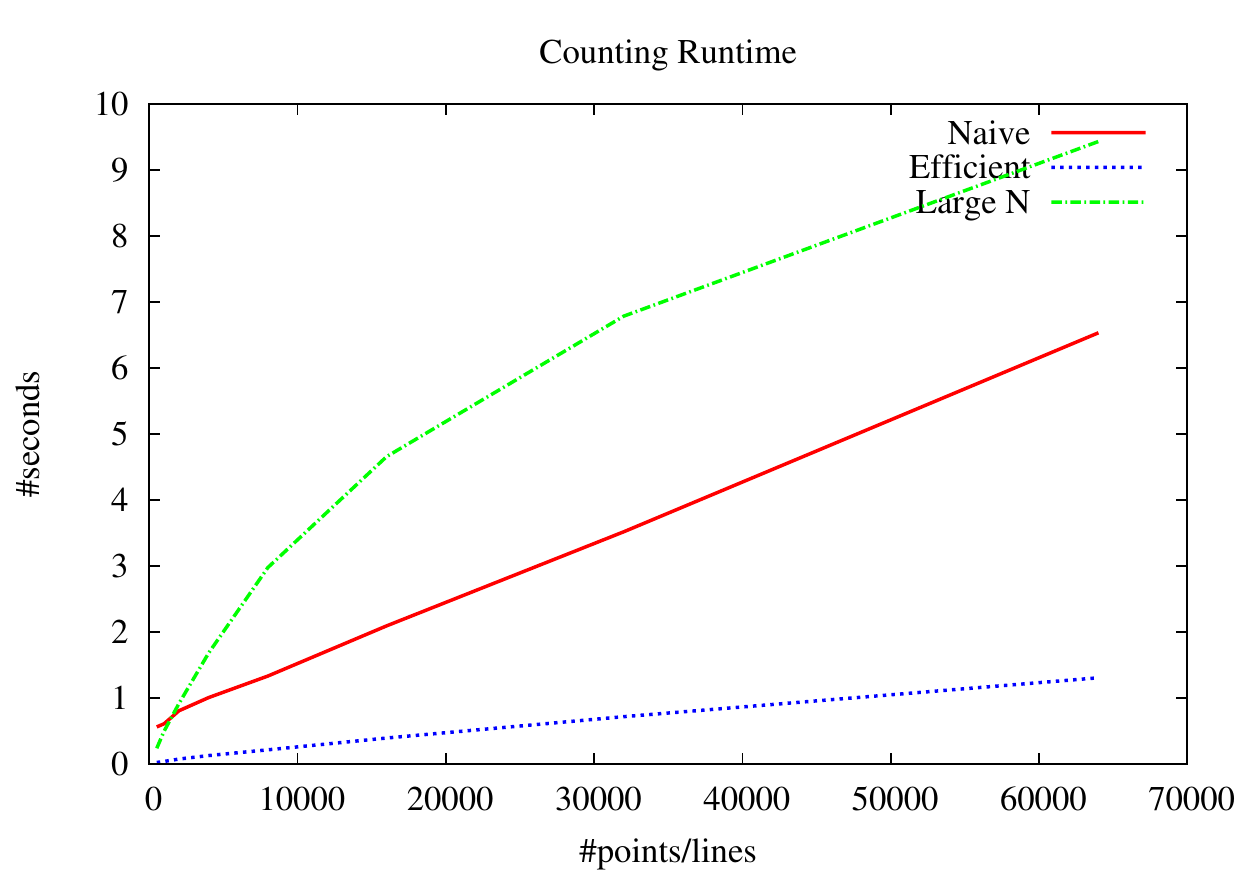}}
  \subcaptionbox{}{\includegraphics[width=0.45\textwidth]{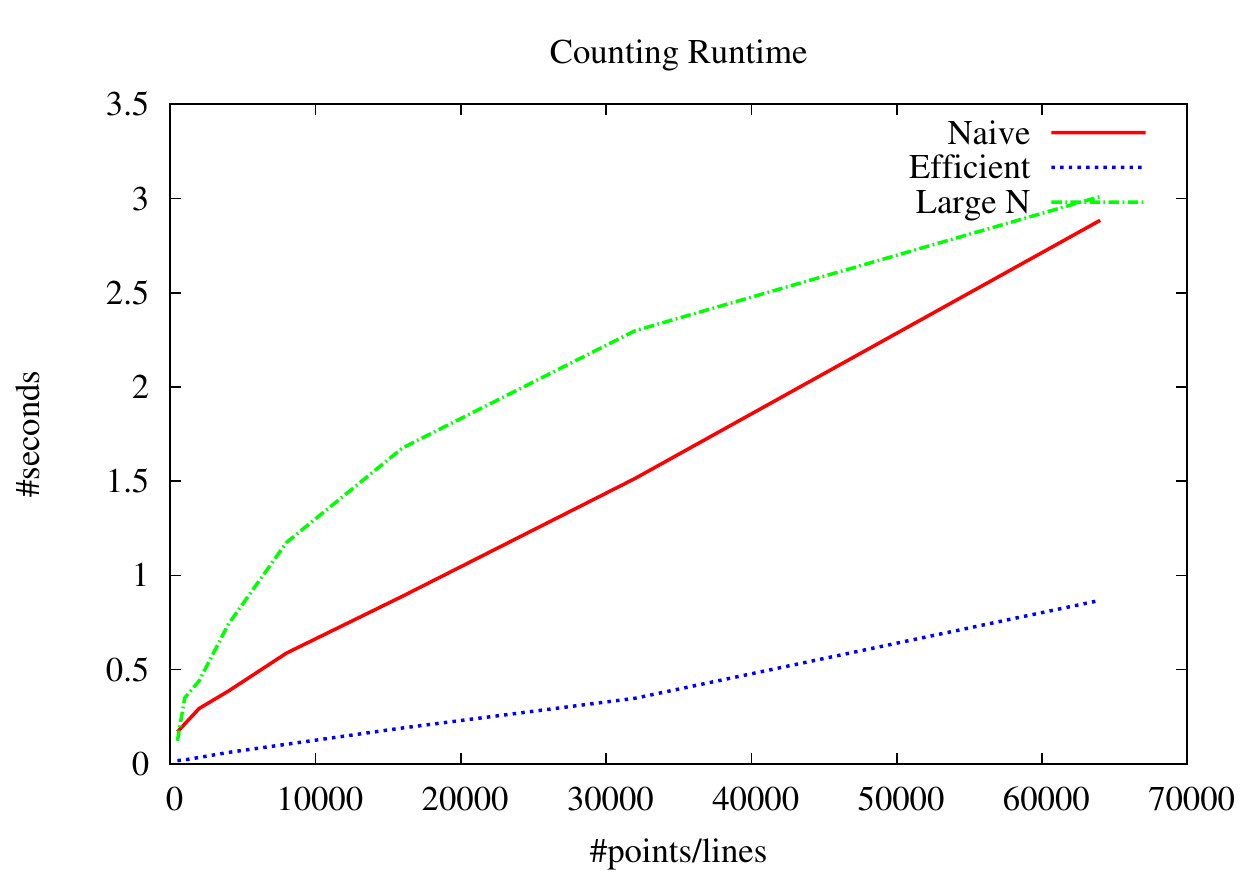}}
  \subcaptionbox{}{\includegraphics[width=0.45\textwidth]{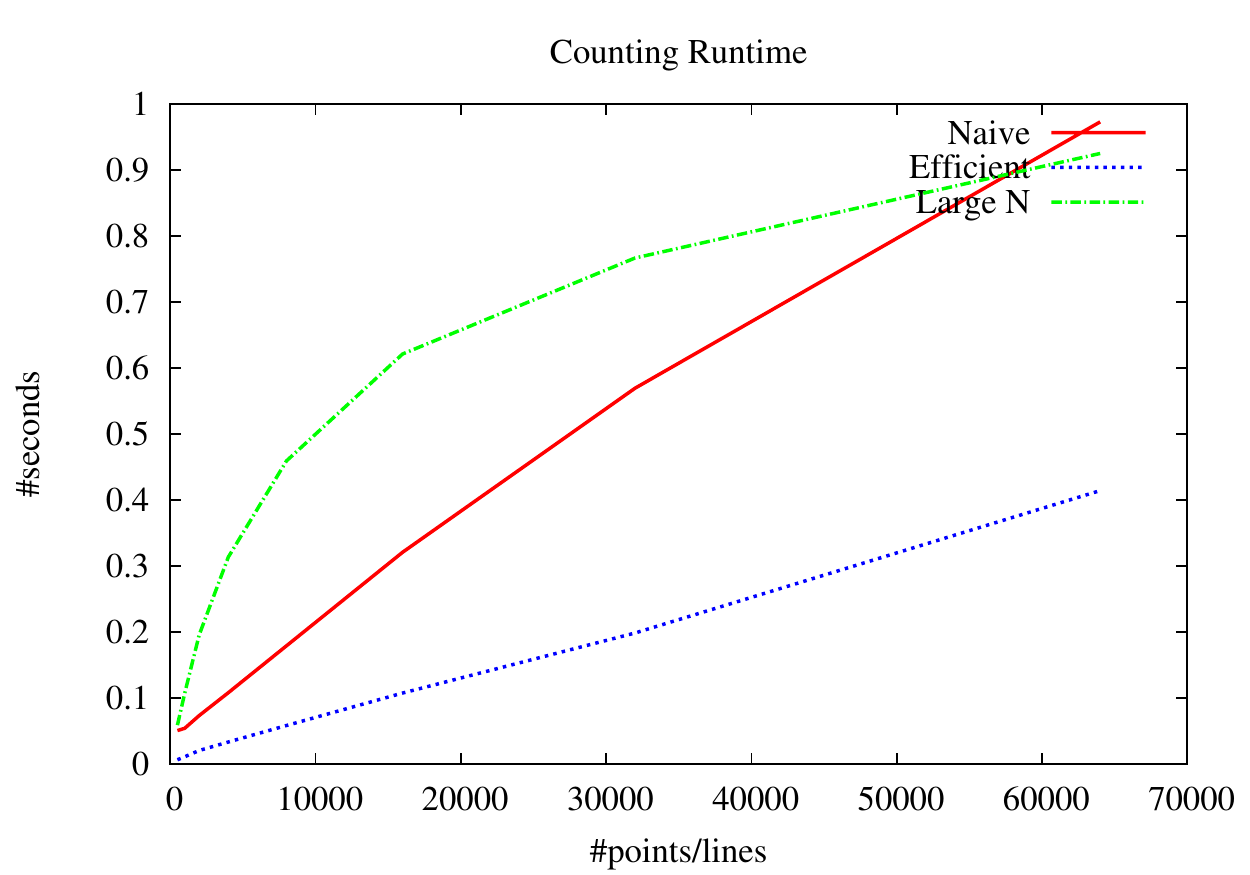}} \hfill

  \caption{The runtime of the counting versions of {\sf Naive}, {\sf Efficient-duality}, and {\sf Large-$n$}
    vs.~the number of points (and lines), for different values of $\eps$:
   (a): $\eps=0.001$; (b): $\eps=0.002$; (c) $\eps=0.004$.}
  \label{fig:same_n}
\end{figure}

\begin{figure}[!htb]
  \centering
  \subcaptionbox{}{\includegraphics[width=0.45\textwidth]{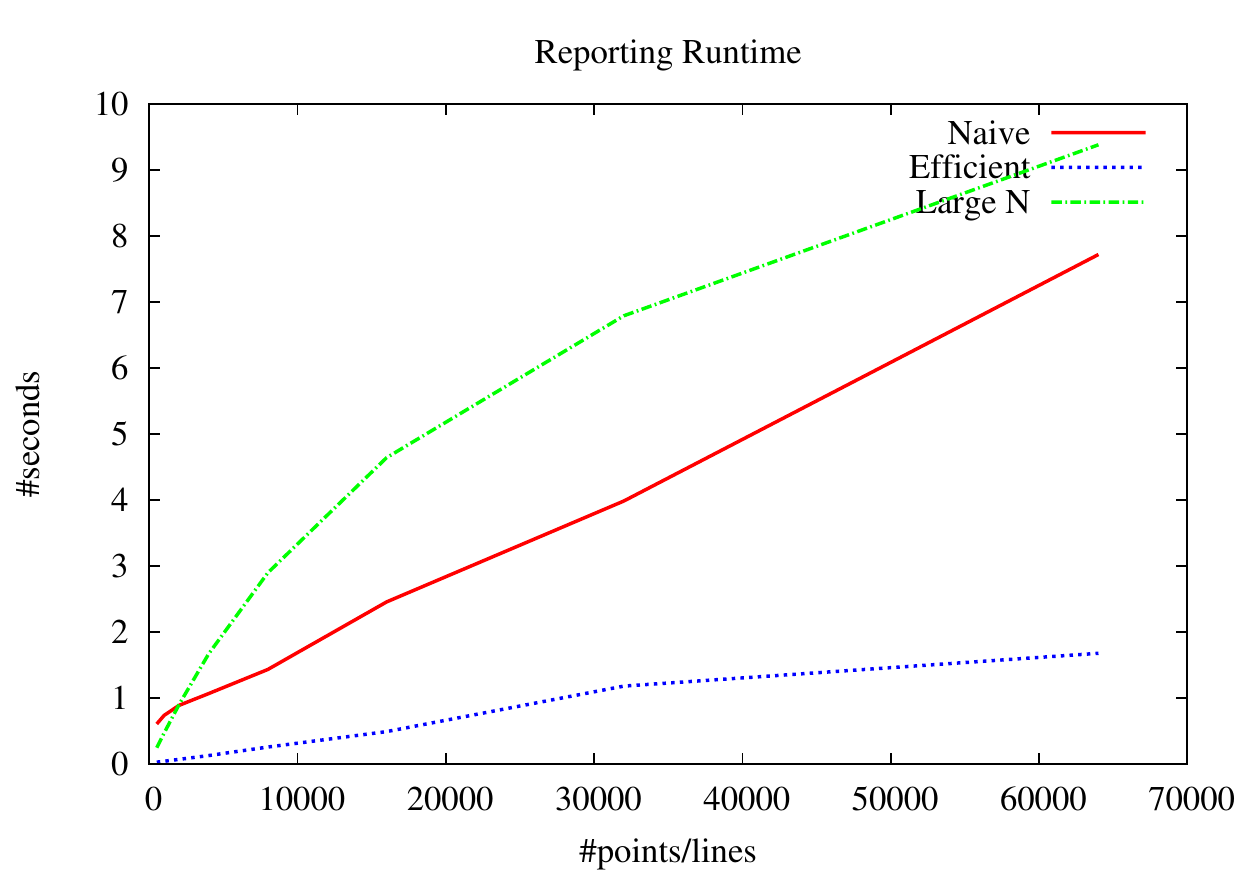}}
  \subcaptionbox{}{\includegraphics[width=0.45\textwidth]{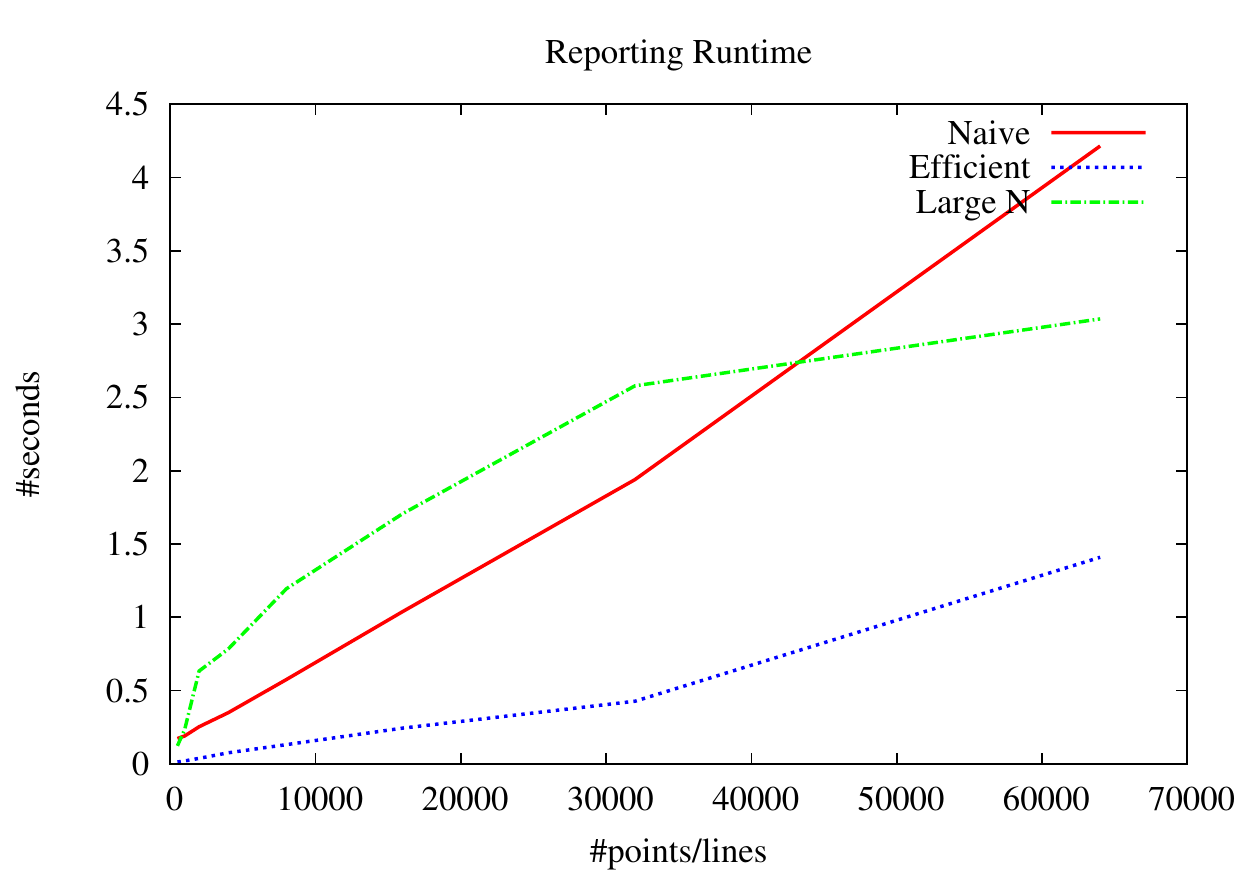}}
  \subcaptionbox{}{\includegraphics[width=0.45\textwidth]{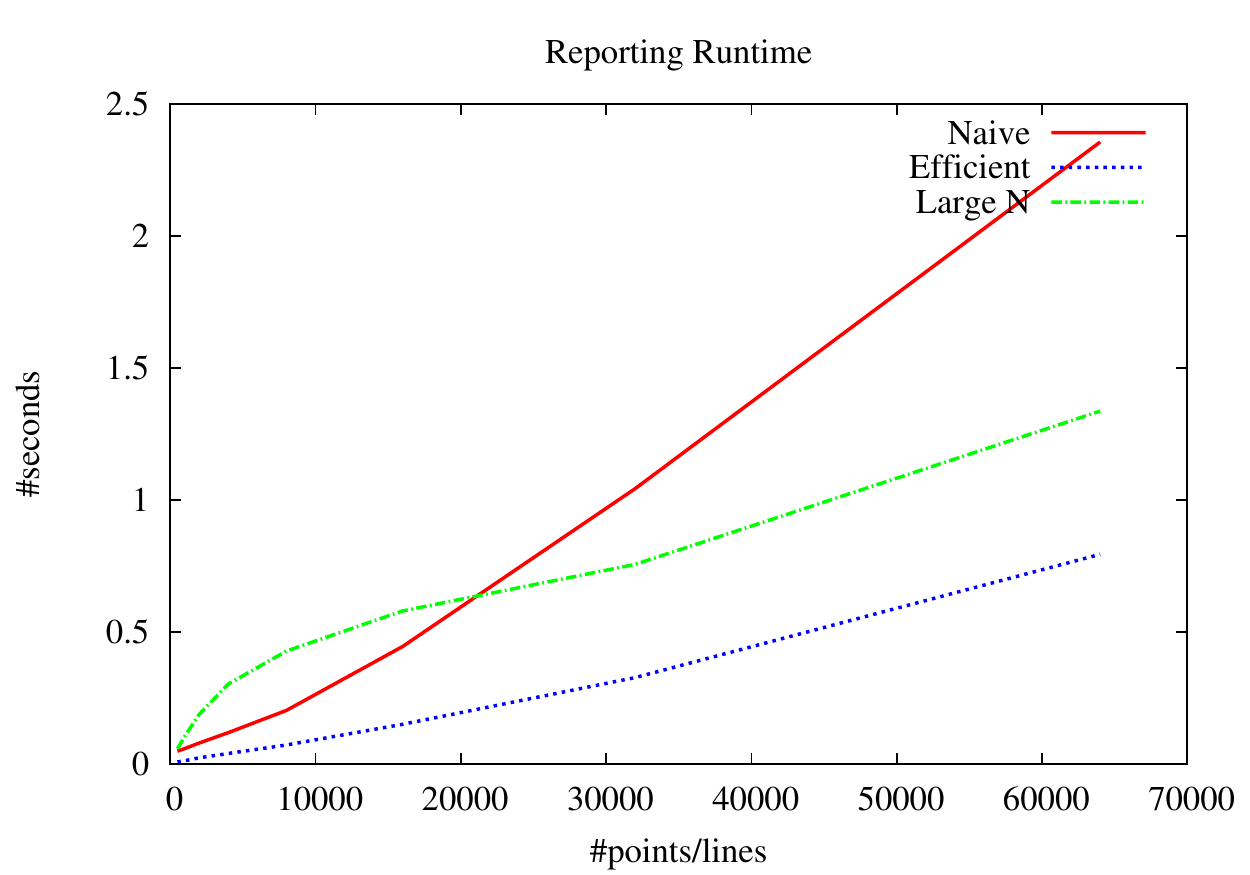}} \quad
  \caption{The runtime of the reporting versions of {\sf Naive}, {\sf Efficient-duality}, and {\sf Large-$n$}
    vs.~the number of points (and lines), for different values of $\eps$:
   (a): $\eps=0.001$; (b): $\eps=0.002$; (c) $\eps=0.004$.}
  \label{fig:same_n_k}
\end{figure}

\medskip
\noindent
{\bf Random points.}
Figures \ref{fig:same_n} and \ref{fig:same_n_k}
show the runtime of the three algorithms {\sf Naive}, {\sf Efficient-duality}, and {\sf Large-$n$} for various values of
$n$ and $\eps$ (since the number of points is the same as the number of lines, there is no need to consider {\sf Naive-duality}).
Each of the three subfigures (a)--(c), in both figures, is for a different choice of $\eps$, which are, respectively,
$0.001$, $0.002$ and $0.004$. The executions reported in Figure \ref{fig:same_n}
only count the number of output (that is, inspected) pairs, essentially making the running time independent of the corresponding value of $k$.
In contrast, the executions reported in Figure \ref{fig:same_n_k} include the cost of reporting the output pairs,
so their running time also depends on $k$.

As can be seen, {\sf Efficient-duality} always performs considerably better than {\sf Naive}, where the difference is
substantial for a wide range of $n$ and $\eps$. The difference is less significant when $\eps$ increases
(also in the counting versions), but {\sf Efficient-duality} still outperforms {\sf Naive}.
Even  the quadratic growth of
 $k$ in the reporting version still leaves our algorithm superior, for the (fairly wide) ranges of $n$
and $\eps$ depicted in the figures.
The  implementation of the {\sf Large-$n$} algorithm is more complex, resulting in a large constant of
proportionality in the overhead, which makes it efficient only for very large values of $n$ (for practical
values of $\eps$).

While serving as a useful testbed for comparing the algorithms, the
random case is not very practical. Moreover, as can be seen in Figure \ref{fig:same_n_k},
the cost of handling the $k$ output pairs (collecting, inspecting and outputting) tends to become
rather large for larger values of $\eps$, and dominates the runtime.

\begin{figure}[!htb]
  \centering
  \subcaptionbox{}{\includegraphics[width=0.45\textwidth]{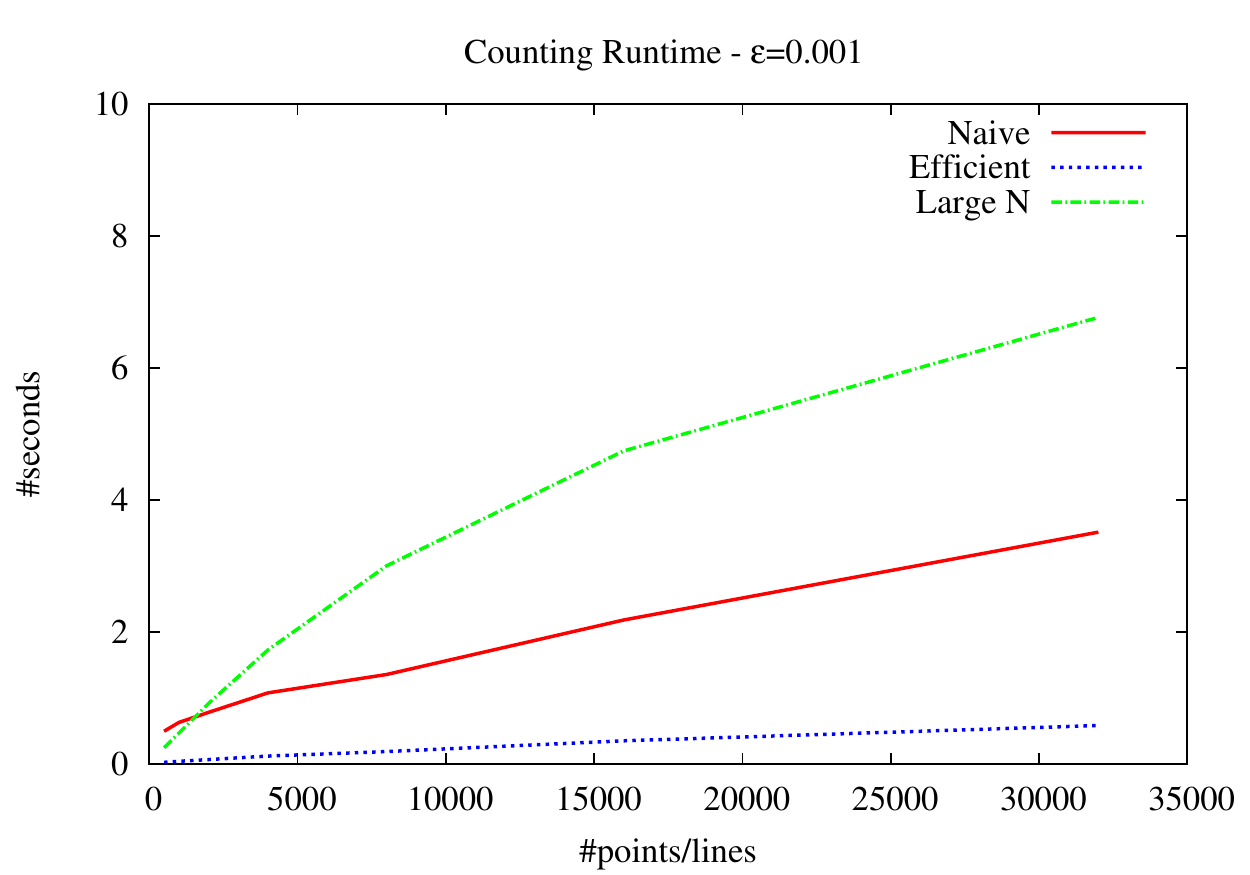}}
  \subcaptionbox{}{\includegraphics[width=0.45\textwidth]{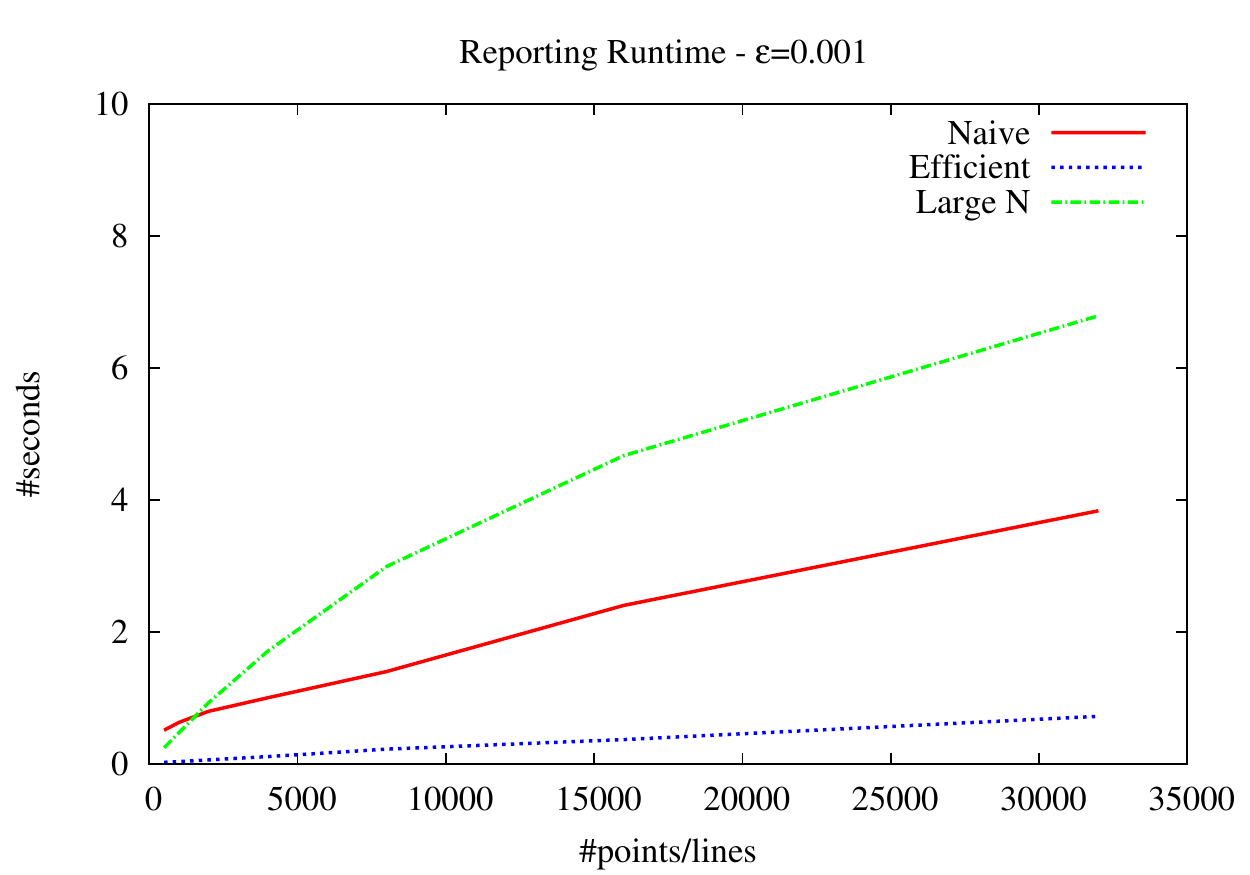}}
  \caption{Image points and lines incidences: (a) Counting (b) Reporting.}
  \label{fig:image_usym}
\end{figure}

\medskip
\noindent
{\bf Real data.}
Our first experiment still used an equal number of points and lines. The results are shown in Figure \ref{fig:image_usym}.
Our algorithm wins with a substantial margin.
Note the rather minor difference between the time for counting
and the time for reporting (because a relatively small number of pairs is reported here).

\begin{figure}[!htb]
  \centering
  \subcaptionbox{}{\includegraphics[width=0.45\textwidth]{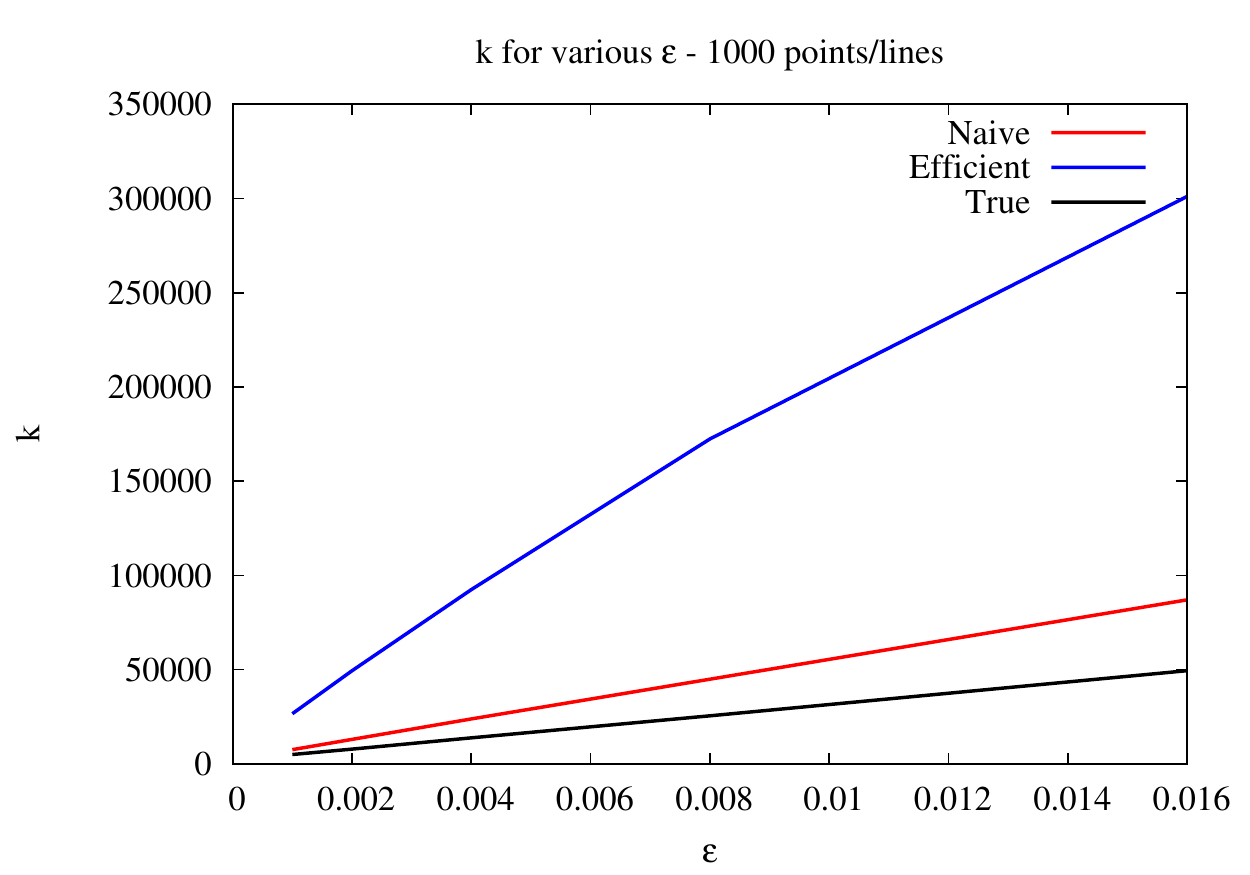}}\quad\quad
  \subcaptionbox{}{\includegraphics[width=0.45\textwidth]{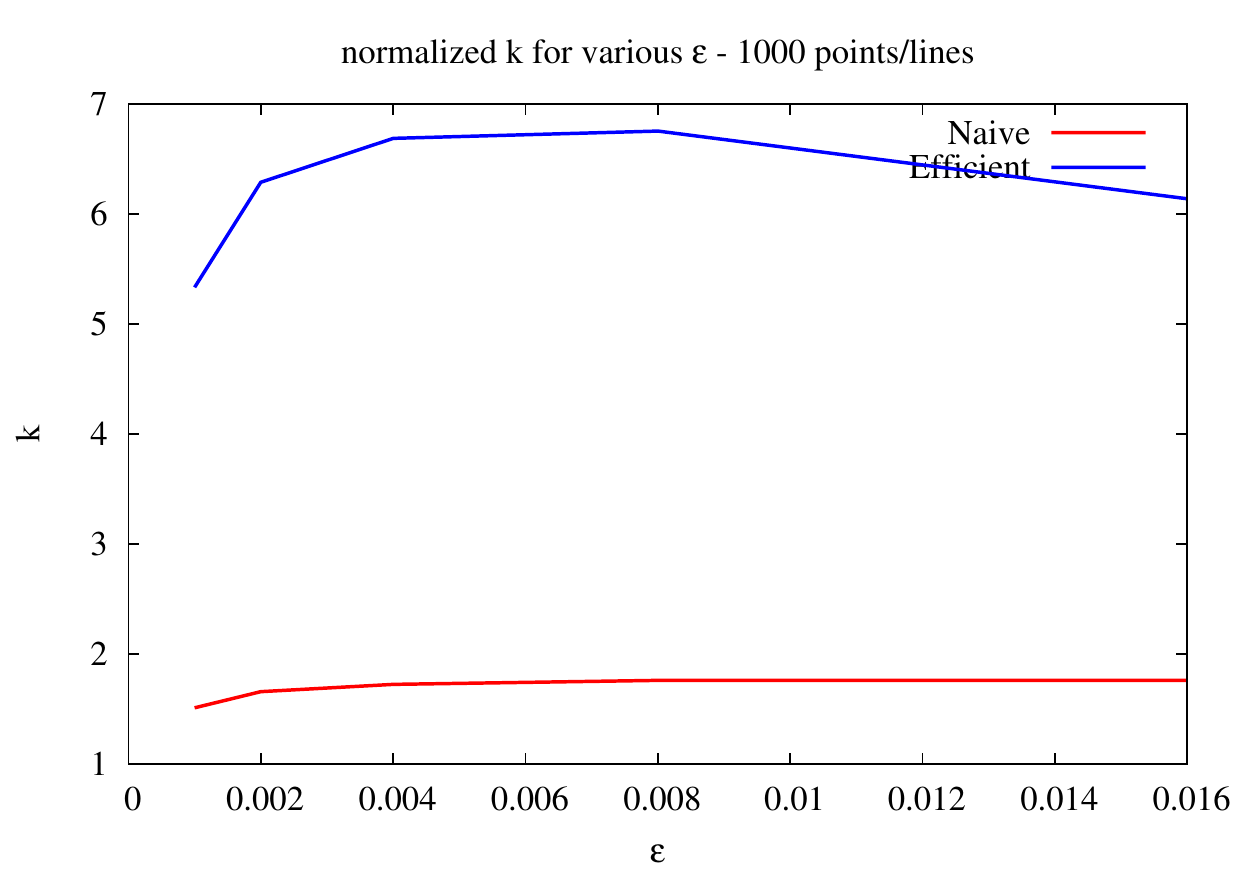}}
  \caption{True pairs and reported pairs by the {\sf Naive} and {\sf Efficient-duality} algorithms.
  (a) The actual number of pairs. (b) The ratio between the number of reported and true pairs.}
  \label{fig:eff_k}
\end{figure}

Figure~\ref{fig:eff_k} shows the actual number of pairs reported  by the {\sf Naive}
and {\sf Efficient-duality} algorithms, as well as the number of true pairs (those at distance at most $\eps$) for various values of $\eps$.
Part (a) shows the actual number of pairs, and part (b) shows the ratio between the numbers of reported and true pairs.
As these figures show, (i) the larger $\alpha$ in the {\sf Efficient-duality} algorithm does indeed causes it to produce more pairs than the
{\sf naive} one; (ii) these numbers grow linearly with $\eps$, as expected; (iii) in fact, the ratio between reported and true pairs
is more or less a constant ($1.8$ for {\sf Naive} and about $5$ for {\sf Efficient-duality}). Still, in spite of this discrepancy
(in favor of {\sf Naive}), the moderate growth of $k$, combined with the much faster overhead, makes our algorithm a clear winner in these
experiments.

\begin{figure}[htb]
  \centering
  \subcaptionbox{}{\includegraphics[width=0.45\textwidth]{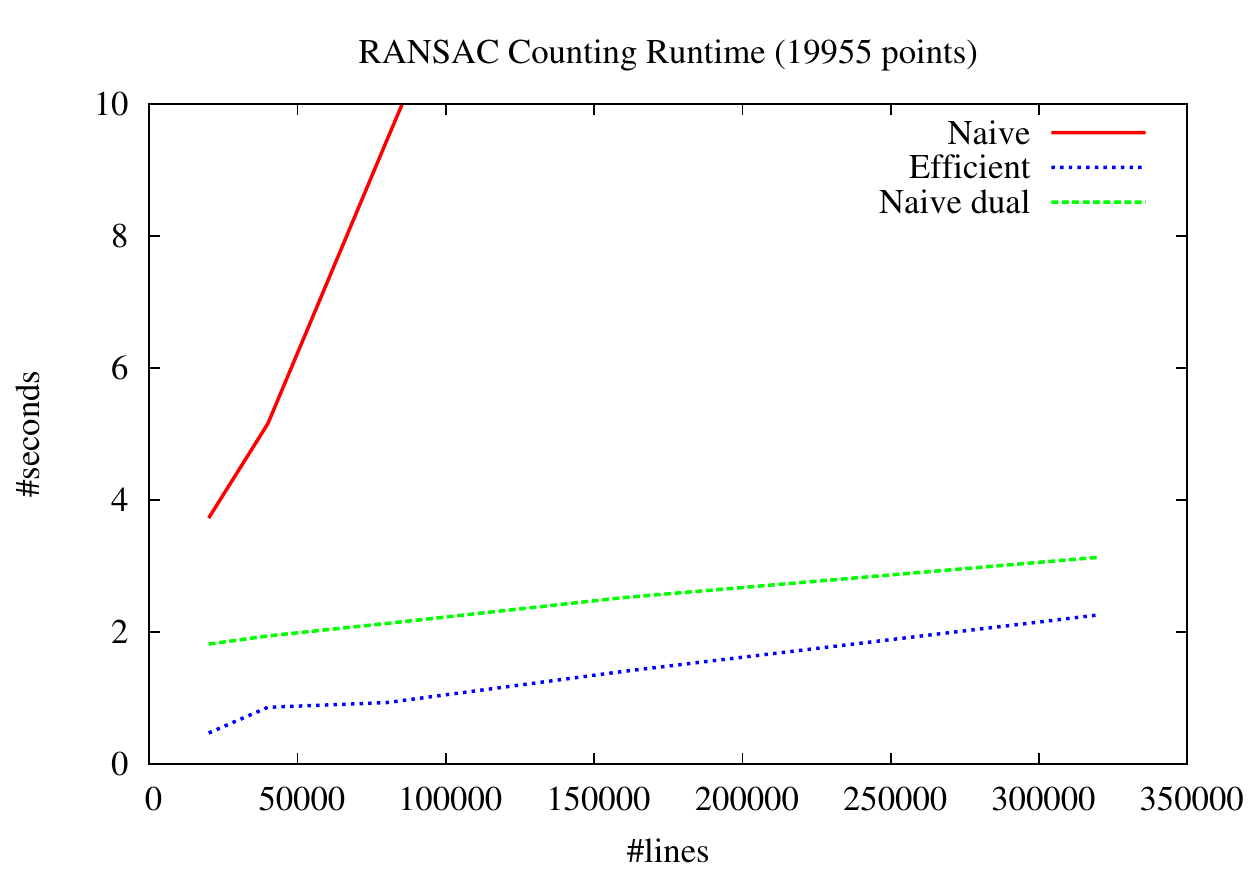}}
  \subcaptionbox{}{\includegraphics[width=0.45\textwidth]{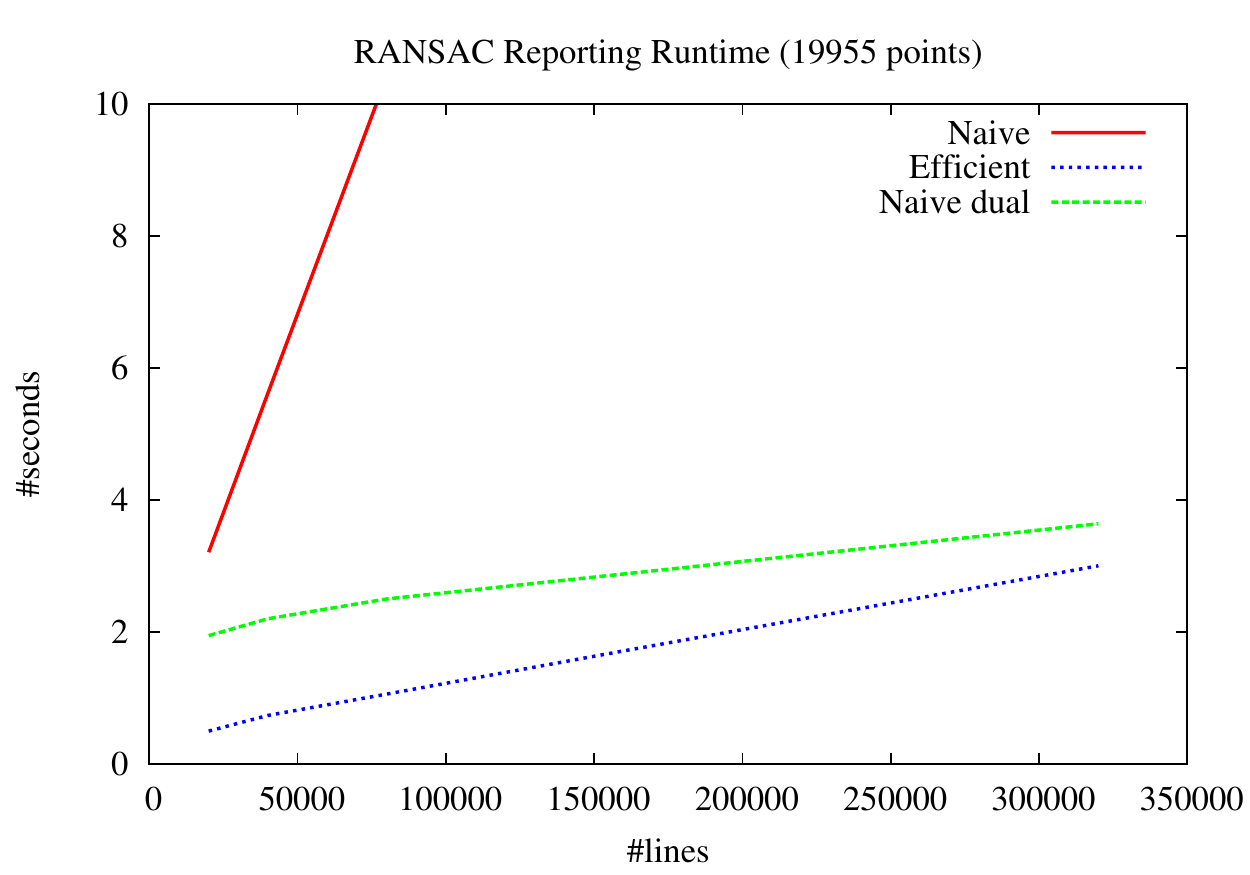}}
  \subcaptionbox{}{\includegraphics[height=3cm, width=0.45\textwidth]{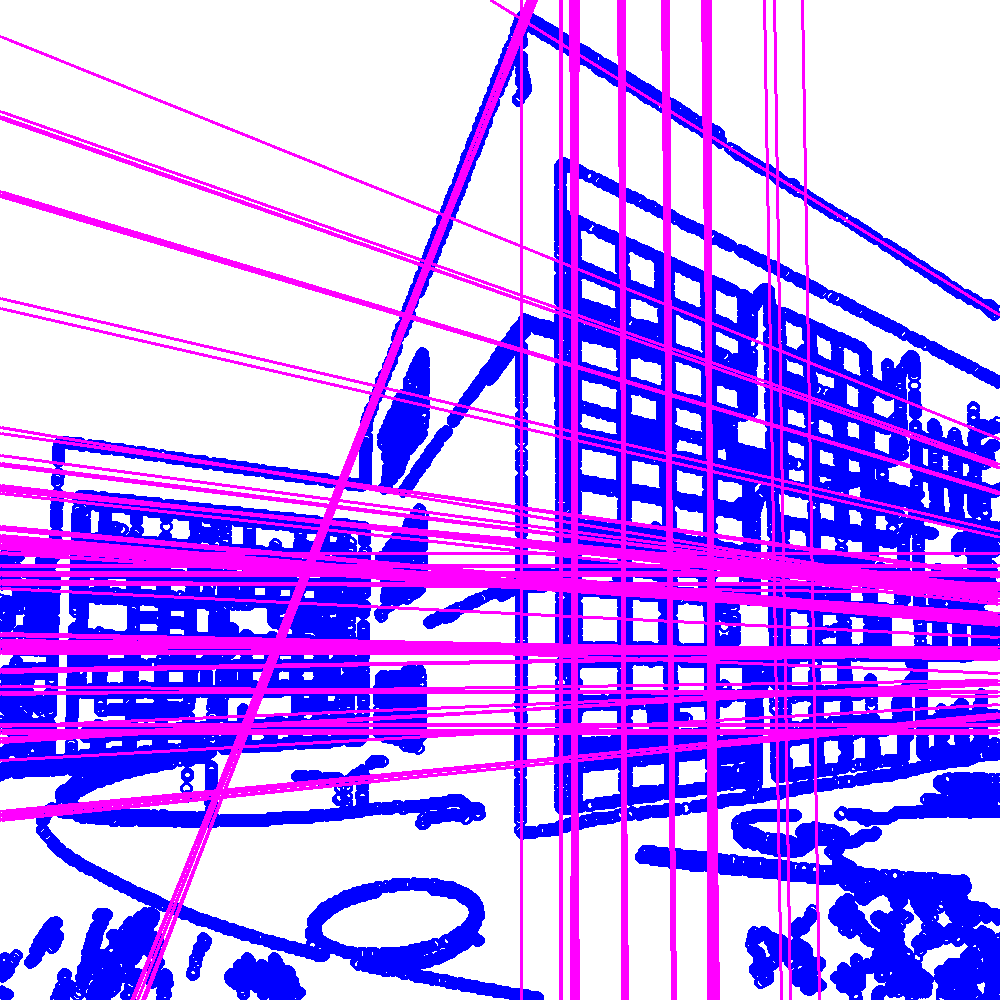}}
  \caption{Line fitting: (a) Counting time. (b) Reporting time. (c) Fitted lines out of 40000 sampled candidate lines. }
  \label{fig:RANSAC}
\end{figure}

\medskip
\noindent
{\bf RANSAC line fitting with our method.}
We ran a complete RANSAC line fitting algorithm where we used the {\sf Naive}, {\sf Naive-duality},
and {\sf Efficient-duality} methods to count and report the nearby points for each candidate line.
The input consists of points sampled from the input image in Figure \ref{fig:images}.
In all experiments we sampled 19955 points, and we randomly generated increasing numbers
of lines by sampling pairs from these points (here the number of lines was not equal to the number of points). Duality allows us to exploit the fact that $n$ and $m$
are different, in both the {\sf Naive-duality} and {\sf Efficient-duality} methods.
The asymptotic theoretical bounds for these two techniques (see the beginning of the section) show that, for a sufficiently large number of lines, namely,
larger than $1/\eps$ times the number of points, {\sf Naive-duality} will become superior to {\sf Efficient-duality}.
Figure \ref{fig:RANSAC} indicates this trend, but shows that the number of lines needed for this to happen (in this example)
has indeed to be very large.

\newpage

\bibliography{esa17-proc-final}{}
\bibliographystyle{plainurl}

\end{document}